\documentclass[preprint,showpacs,preprintnumbers,amsmath,amssymb,superscriptaddress,10pt,graphicx]{revtex4}

\usepackage{graphicx}
\usepackage{dcolumn}
\usepackage{bm}
\usepackage{amssymb}

\begin{document}

\title{Stable indications of relic gravitational waves in Wilkinson Microwave 
Anisotropy Probe data and forecasts for the Planck mission}

\author{W.~Zhao}
\email{Wen.Zhao@astro.cf.ac.uk} \affiliation{School of Physics
and Astronomy, Cardiff University, Cardiff, CF24 3AA, United
Kingdom} \affiliation{Wales Institute of Mathematical and
Computational Sciences, Swansea, SA2 8PP, United Kingdom} 
\affiliation{Department of
Physics, Zhejiang University of Technology, Hangzhou, 310014,
People's Republic of China}

\author{D.~Baskaran}
\email{Deepak.Baskaran@astro.cf.ac.uk} \affiliation{School of Physics
and Astronomy, Cardiff University, Cardiff, CF24 3AA, United
Kingdom} \affiliation{Wales Institute of Mathematical and
Computational Sciences, Swansea, SA2 8PP, United Kingdom}

\author{L.P.~Grishchuk}
\email{Leonid.Grishchuk@astro.cf.ac.uk} \affiliation{School of Physics
and Astronomy, Cardiff University, Cardiff, CF24 3AA, United
Kingdom} \affiliation{Sternberg Astronomical Institute, Moscow State University, Moscow, 
119899, Russia}



\begin{abstract}
{\small  The relic gravitational waves are the cleanest probe of the violent
times in the very early history of the Universe. They are expected to leave
signatures in the observed cosmic microwave background anisotropies. We 
significantly improved our previous analysis \cite{zbg} of the 5-year WMAP 
$TT$ and $TE$ data at lower multipoles $\ell$. This more general analysis
returned essentially the same maximum likelihood (ML) result (unfortunately, 
surrounded by large remaining uncertainties): the relic gravitational waves 
are present and they are responsible for approximately $20\%$ of the temperature 
quadrupole. We identify and discuss the reasons by which the contribution of 
gravitational waves can be overlooked in a data analysis. One of the reasons
is a misleading reliance on data from very high multipoles $\ell$, another - 
a too narrow understanding of the problem as the search for $B$-modes of 
polarization, rather than the detection of relic gravitational waves with the 
help of all correlation functions. Our analysis of WMAP5 data has led to the
identification of a whole family of models characterized by relatively high 
values of the likelihood function. Using the Fisher matrix formalism we 
formulated forecasts for {\it Planck} mission in the context of this family of
models. We explore in details various `optimistic', `pessimistic' and `dream 
case' scenarios. We show that in some circumstances the $B$-mode detection 
may be very inconclusive, at the level of signal-to-noise ratio $S/N =1.75$, 
whereas a smarter data analysis can reveal the same gravitational wave signal 
at $S/N= 6.48$. The final result is encouraging. Even under unfavourable 
conditions in terms of instrumental noises and foregrounds, the relic 
gravitational waves, if they are characterized by the ML
parameters that we found from WMAP5 data, will be detected by {\it Planck} at 
the level $S/N = 3.65$.}

\end{abstract}


\maketitle

\section{Introduction \label{section1}}

A complete cosmological theory is supposed to explain not only the 
present state of the observed Universe (see, for example, \cite{cosm}), but 
also its early dynamical behaviour and possibly its birth \cite{zg}. Our 
present state $P$ is characterized by approximate large-scale homogeneity and 
isotropy  within a patch of the size 
$l_p \approx 10^{3} l_H \approx 10^{31}{\rm cm}$ (see last paper in \cite{zg})
and the averaged energy density $\rho_{p}c^2$ 
of all sorts of matter in this patch $\rho_p =3 H_0^2/8 \pi G 
\approx 10^{-29} {\rm g}/{\rm cm}^3$, where $H_0$ is the present day Hubble 
parameter and $l_H = c/H_0 \approx 10^{28}{\rm cm}$. The limits of applicability of the 
currently available theories are set by the Planck density 
$\rho_{\rm Pl} = c^5/G^2 \hbar \approx 10^{94}{\rm g}/{\rm cm}^{3}$ 
and the Planck size $l_{\rm Pl} = (G \hbar/c^3)^{1/2} \approx 10^{-33}{\rm cm}$.
One can imagine that the embryo Universe was created by a quantum-gravity 
(or by a `theory-of-everything') process. The emerging classical configuration 
was probably characterized by the near-Planckian energy density and size.
The total energy, including gravity, was likely to be zero then, 
and remains zero now. 

The problem is that this hypothesis requires further assumptions. The arising
classical configuration can not reach the present 
state $P$ if it expands all the time according to the usual laws 
of radiation-dominated and matter-dominated evolution. By the time the 
Universe (i.e. the patch of approximate homogeneity and isotropy) has reached
the size $l_p$, its energy density would have dropped to the level many orders 
of magnitude lower than the required $\rho_p$. Therefore, the newly born Universe needs 
a primordial kick before it can join the pathway of normal radiation-dominated 
expansion. The kick should allow the size of the patch to increase by about 
33 orders of magnitude without losing too much of the energy density of 
whatever substance that was there, or maybe even slighly increasing this 
energy density at the expense of the energy density of the gravitational field.

The relic gravitational waves \cite{gr74} are necessarily generated by a 
strong variable gravitational field of the very early Universe. They are the 
cleanest probe of what was happening during the violent times of the
initial kick. Specifically, the quantum-mechanical Schr\"odinger 
evolution transformes the initial no-particle (vacuum) state of the 
gravitational waves into a multiparticle (strongly squeezed vacuum) state. 
Under certain additional conditions, the same holds true for other degrees 
of freedom of the gravitational field (metric) perturbations, including those 
representing the density perurbations. As a result, the patch of homogeneity 
and isotropy will necessarily be augmented by primordial cosmological 
perturbations of quantum-mechanical origin. This process is called the 
superadiabatic, or parametric, amplification; for a recent review of the 
subject, see the last paper in \cite{gr74}. 
	
As before (see \cite{zbg} and references there), we are working with	
\begin{eqnarray}
 ds^2=-c^2dt^2+a^2(t)(\delta_{ij}+h_{ij})dx^idx^j 
 =a^2(\eta)[-d\eta^2+(\delta_{ij}+h_{ij})dx^idx^j]
\nonumber
\label{metric}
\end{eqnarray}
and the Fourier-expanded gravitational field (metric) perturbations
\begin{eqnarray}
\label{h}
 h_{ij}(\eta,{\bf
 x})=\frac{\mathcal{C}}{(2\pi)^{3/2}}\int_{-\infty}^{+\infty}\frac{d^3{\bf
 n}}{\sqrt{2n}} \sum_{s=1,2}\left[\stackrel{s}{p}_{ij}({\bf n})
 \stackrel{s}{h}_{n}(\eta)e^{ i{\bf n}\cdot{\bf x}}
\stackrel{s}{c}_{\bf n}+ \stackrel{s~*}{p_{ij}}({\bf n})
\stackrel{s~*}{h_{n}}(\eta)e^{-i{\bf n}\cdot{\bf x}}
\stackrel{s~\dag}{c_{\bf n}} \right].
\end{eqnarray}
The polarization tensors $\stackrel{s}{p}_{ij}({\bf n})$ ($s=1,2$)
describe either the two transverse-traceless components of 
gravitational waves (gw), or the scalar and longitudinal-longitudinal
components of density perturbations (dp). Assuming the initial vacuum state of
participating perturbations, the resulting metric power spectra $h^2(n,\eta)$
are given by 
\begin{eqnarray}
\langle 0| h_{ij}(\eta,{\bf x})h^{ij}(\eta,{\bf x}) |0 \rangle
= \int\limits_{0}^{\infty} \frac{dn}{n}~h^2(n,\eta),~~~~~
\label{powerspectrumhdef}
h^2(n,\eta) \equiv \frac{\mathcal{C}^2}{2\pi^2}n^2
\sum_{s=1,2}|\stackrel{s}{h}_n(\eta)|^2,
\end{eqnarray} 
where the mode functions $\stackrel{s}{h}_{n}(\eta)$ are taken either from gw 
or dp equations, and $\mathcal{C}=\sqrt{16\pi}l_{\rm Pl}$ for gravitational
waves and $\mathcal{C}=\sqrt{24\pi}l_{\rm Pl}$ for density perturbations.

The numerical levels and shapes of the generated power spectra are 
determined by the strength and variability of the gravitational `pump' 
field. The simplest assumption about the initial kick is that its entire 
duration can be described by a single power-law scale factor \cite{gr74}
\begin{equation}
\label{inscf}
a(\eta) = l_o|\eta|^{1+ \beta},
\end{equation}
where $l_o$ and $\beta$ are constants, $\beta < -1$. Then, the generated 
primordial metric power spectra (for wavelengths longer than the Hubble 
radius at that time) have the universal power-law dependence on the 
wavenumber $n$:
\begin{equation}
\label{primsp}
h^{2}(n) \propto n^{2(\beta+2)}.
\end{equation}

It is common to write these metric power spectra separately for gw and dp:
\begin{equation}
\label{primsp}
h^2(n)~({\rm gw}) = B_t^2 n^{n_t}, ~~~~~h^2(n)~({\rm dp})=B_s^2 n^{n_s -1}.
\end{equation}
In the case of power-law scale factors (\ref{inscf}) 
(or piece-wise power-law scale factors), the equations for 
metric perturbations representing gravitational waves and density 
perturbations are exactly the same \cite{Grishchuk1994}. Therefore, 
according to the theory of quantum mechanical generation of cosmological 
perturbations \cite{Grishchuk1994}, the spectral indices 
are approximately equal, $n_s-1 = n_t = 2(\beta+2)$, and the 
amplitudes $B_t, B_s$ are of the order of magnitude of the ratio 
$H_i/H_{\rm Pl}$, where $H_i\sim c/l_o$ is the characteristic value of the 
Hubble parameter during the kick. (An initial kick driven by a scalar field 
is usually associated with inflation.) In what follows, 
we are using the numerical code CAMB \cite{CAMB} 
and related notations for gw and dp power spectra adopted there:
\begin{eqnarray}
\label{PsPt}
P_{t}(k)=A_t(k/k_0)^{n_t},~~P_{s}(k)=A_s(k/k_0)^{n_s-1},
\end{eqnarray}
where $k_0=0.002$Mpc$^{-1}$.

There is no doubt that the metric perturbations with wavelengths greater 
than the Hubble radius in the times of recombination do exist. Indeed, it is 
known for long time \cite{gz78} that it is precisely this sort of 
long-wavelength metric perturbations that provide the main contribution 
to the lower multipoles, starting from $\ell =2$, of the Cosmic Microwave 
Background (CMB) temperature anisotropies. The very existence of CMB 
anisotropies at lower $\ell$'s \cite{smoot,wmap5} testifies to the existence 
of such long-wavelength perturbations. They are likely to be the perturbations 
of quantum-mechanical origin. 

The assumption of a single power-law index $\beta$ is the simplest and 
easiest to analyze, but it is too strong. It has the consequence that one 
and the same spectral index describes the interval of 30 orders of magnitude 
of wavelengths in the primordial power spectra. In reality, as it appears from 
our CMB analysis below, even at the span of 2 orders of magnitude in terms of 
wavelengths the spectral index $n_s$ is likely to be somewhat different. 
We will discuss this point in more detail in the text of this paper. 

We start (Sec.~\ref{section2}) with a significant improvement of our previous 
analysis \cite{zbg} of the 5-year Wilkinson Microwave Anisotropy Probe (WMAP5) 
$TE$ data at $\ell \leq 100$  \cite{LAMBDA}. In contrast to \cite{zbg}, we 
work directly with both $TT$ and $TE$ 
datasets and impose no restrictions on the (constant) perturbation parameters 
$A_t, A_s, n_t, n_s$, except $n_s - 1 = n_t$ which is implied by the theory of 
quantum-mechanical generation. We work with the quadrupole ratio $R$ 
\begin{eqnarray}
\label{defineR}
 R \equiv \frac{C_{\ell=2}^{TT}({\rm gw})}{C_{\ell=2}^{TT}({\rm dp})}
\end{eqnarray}
and the remaining two free parameters $A_s, n_s$. This more general analysis 
returns essentially the same as before \cite{zbg} maximum likelihood (ML) 
values: $R=0.229$ and $n_s=1.086$, with approximately the same as before 
uncertainties. We demonstrate that the data at the multipole $\ell=2$ 
(dubbed ``anomalously low" in the literature) are not 
to be blamed for these determinations. After removal of this data point 
altogether, the ML values do not change much. These improvements and 
cross-checks make more stable and robust our conclusion \cite{zbg} that the
WMAP5 data do contain a hint of presence of relic gravitational waves.    

In Sec.~\ref{section3} and Sec.~\ref{section4} we show in detail how relic 
gravitational waves can be overlooked in CMB data analysis. In 
Sec.~\ref{section3} we concentrate on one of the reasons, which is the 
attempt of placing the ever ``tighter" constraints on gravitational
waves by using the data from high $\ell$'s of CMB and large-scale structure
surveys. These data have nothing to do with gravitational waves and they 
can mislead the identification of gw contribution. We show that even the use 
of CMB data from the adjacent interval $101 \le \ell \le 220$, where the 
role of gravitational waves is already small, is dangerous. We defer to 
Sec.~\ref{section4} a detailed discussion of another recipe to overlook 
the relic gravitational waves. This is the wide-spread `obsession' with the 
detection of $B$-modes, rather than the detection of relic gravitational 
waves with the help of all available observational CMB channels.

The prospects of observing relic gravitational waves by the 
already deployed {\it Planck} satellite \cite{planck} are analysed in great 
detail in Sec.~\ref{section4}. We adopt improved evaluations of 
foregrounds \cite{remove1}, \cite{foreground3}, \cite{cmbpol} 
and instrumental noises. The main thrust of the section is the comparison of 
the performances of various combinations of observational channels:
$TT+TE+EE+BB$, $TT+TE+EE$, and $BB$ alone. We discuss different models of the 
foregrounds and their subtraction, individual `optimistic', `pessimistic'
and `dream case' scenarios, as well as complications in the data analysis 
itself. The final conclusions are formulated not only for the model 
characterized by the set 
of ML parameters derived from the WMAP5 data, but also for the whole 
class of models characterized by the high values of the 3-parameter 
likelihood function. We show that there exists plenty of situations where the 
results from the $BB$ channel alone are inconclusive, whereas a smarter data 
analysis can reveal a significant detection. For other approaches to observing 
relic gravitational waves in the CMB temperature and polarization anisotropies 
see \cite{PolnarevKeating,Pagano2007}.

The good news is that even under unfavorable conditions, the {\it Planck} 
satellite will see the relic gw signal (assuming that it has the WMAP5 
maximum likelihood value $R=0.229$) at a better than 3$\sigma$ level.  
Furthermore, we believe that the methods and evaluations of this paper
can also be used in ground-based and balloon-borne experiments 
\cite{BICEP,quad,Clover,QUITE,EBEX,SPIDER,keating2009}.

At the end of the introduction, it is important to stress that the current 
thinking in this area of science is greatly influenced by inflationary 
understanding of quantum mechanics and relativity: ``Quantum fluctuations, 
usually observed only on microscopic scales, were stretched to 
astronomical sizes and promoted to cosmic significance as the seeds 
of large scale structure" \cite{dod}, ``the superluminal expansion of space 
during inflation stretched these scales outside
of the horizon" \cite{baumann}, ``Inflation...stretched space...and promoted 
microscopic quantum fluctuations to perturbations on cosmological scales. 
Inflation makes detailed predictions..." \cite{EPIC}, and so on. 
	
Indeed, inflationary views on physics have translated into inflationary 
observational predictions. They are encapsulated in the formula 
for the predicted scalar metric power spectrum of density perturbations 
$P_s \approx P_t/\epsilon$, which is divergent at small $\epsilon$ ($\epsilon 
\equiv -{\dot {H}}/H^2$, $\epsilon =0$ in the standard inflation), and the 
detailed `tensor-to-scalar' ratio $r$ ($r \equiv P_t/P_s$): 
\begin{equation}
\label{tsr}
r= 16 \epsilon = - 8 n_t.
\end{equation}
	
The widely quoted limits on $r$, $r<0.22 $ ($95\%$ C.L.) \cite{wmap5} 
were derived from the likelihood function for $r$. The analysis 
has resulted in the maximum likelihood value $r_{ML}=0$. Since 
$P_t \approx (H_i/H_{\rm Pl})^2 \neq 0$ and $r_{ML}=0$, 
one has to decide whether the most likely values of density 
perturbations $P_s$ responsible for the data collected and analyzed by the 
WMAP team are infinitely large, or inflationary predictions are wrong. 
The existing and planned data analyses are usually based on the enforced 
(incorrect) inflationary relation $r= - 8n_t$; the final physical conclusions 
are formulated in terms of constraints imposed on the (possibly non-existent) 
scalar field, and so on (see, for example, \cite{wmap5}, \cite{EG}). 
	
As for the quantity $r$, there is no doubt that, in general, the 
inflationary theory can predict for $r$ everything what one can 
possibly ask for (for a review, see \cite{baumann} and references there). 
But the most advanced inflationary theories, operating with 
{\it warped D-brane inflation} \cite{kachru}, \cite{baummc}, 
{\it D3-brane inflation in warped throats} \cite{st}, {\it string theory 
inflation} \cite{kl}, etc., either ``allow a very low tensor amplitude 
$r\ll10^{-4}$", or lead to the conclusion that ``D3-brane inflation in 
Calabi-Yau throats, or in most tori, cannot give rise to an observably-large 
primordial tensor signal", or to the conclusion that $r \approx 10^{-24}$ and 
the ``existing models of string theory inflation do not predict a detectable 
level of tensor modes". These conclusions make the search for the 
{\it inflationary} gravitational waves (i.e. relic gravitational waves as 
presented by inflationists) a senseless enterprise. 
	
Obviously, in this paper, we are not using the inflationary theory and 
its observational predictions. (These predictions are based on the 
inflationary hat-trick of extracting arbitrarily large scalar metric 
perturbations $P_s$ out of vacuum fluctuations of the scalar field.
For a more detailed criticism of inflationary theory, see last papers 
in \cite{gr74} and \cite{zg}.) 


\section{Improved evaluation of relic gravitational waves from WMAP $TT$ 
and $TE$ data \label{section2}}


\subsection{Likelihood functions and summary of the previous results 
\label{section2.1}}

Relic gravitational waves compete with density perturbations in 
generating CMB temperature and polarization anisotropies  
at relatively low multipoles $\ell\lesssim100$. 
For this reason we focus on the WMAP data at $\ell\le100$.
As before \cite{zbg}, we use the symbols $C_{\ell}^{TT}$, $C_{\ell}^{TE}$, 
$C_{\ell}^{EE}$, $C_{\ell}^{BB}$ for CMB power spectra and 
$D_{\ell}^{TT}$, $D_{\ell}^{TE}$, $D_{\ell}^{EE}$, $D_{\ell}^{BB}$ for their 
estimators. In this section we ignore the $B$-mode of polarization because WMAP puts only 
upper limits on it. 

The variables $D_{\ell}^{TT}$, $D_{\ell}^{TE}$ and $D_{\ell}^{EE}$ obey the 
Wishart probability density function (pdf) 
\cite{zbg,wishart3,wishart1,wishart2}
\begin{eqnarray}\label{wishart}
f(D_{\ell}^{TT},D_{\ell}^{TE},D_{\ell}^{EE})&=&
\left\{\frac{1}{4(1-{\rho_{\ell}^2})({\sigma_\ell^T}{\sigma_\ell^E})^2}
\right\}^{n/2} \frac{n^3({x}{y}-{z}^2)^{(n-3)/2}}{\pi^{1/2}\Gamma(n/2)
\Gamma((n-1)/2)} \nonumber\\
&\times&
\exp\left\{-\frac{1}{2(1-{\rho_\ell^2})}\left(\frac{{x}}{(\sigma_\ell^T)^2}+
\frac{{y}}{(\sigma_\ell^E)^2}-\frac{2{\rho_l}
{z}}{{\sigma_\ell^T}{\sigma_\ell^E}}\right)\right\},
\end{eqnarray}
where $f_{\rm sky}$ is the sky-cut factor, $f_{\rm sky}=0.85$ for WMAP and 
$f_{\rm sky}=0.65$ for {\it Planck}, and $n= (2\ell+1)f_{\rm sky}$ is the 
number of effective degrees of freedom at multipole $\ell$. $\Gamma$ is the 
$Gamma$-function. This pdf contains the variables $D_{\ell}^{XX'}$ 
($XX'=TT,TE,EE$) in quantities ${x}\equiv n(D_\ell^{TT}+N_{\ell}^{TT})$, 
${y}\equiv n(D_\ell^{EE}+N_{\ell}^{EE})$, ${z}\equiv nD_\ell^{TE}$. The 
information on the power spectra $C_{\ell}^{XX'}$ is contained in quantities 
\begin{eqnarray}
\label{rho}\nonumber
{\sigma_\ell^T}= \sqrt{C_\ell^{TT}+N_\ell^{TT}}, ~\sigma_\ell^E=
\sqrt{C_\ell^{EE}+N_{\ell}^{EE}},~
{\rho_\ell}=\frac{C_\ell^{TE}}{\sqrt{(C_\ell^{TT}+N_{\ell}^{TT})(C_\ell^{EE}+
N_{\ell}^{EE})}},
\end{eqnarray}
where $N_{\ell}^{TT}$ and $N_{\ell}^{EE}$ are the total noise power spectra.

We are mostly interested in $TT$ and $TE$ data, so we shall work with the 
joint pdf for $D_{\ell}^{TT}$ and $D_{\ell}^{TE}$. This pdf is derived from 
(\ref{wishart}) by integrating over the variable $D_{\ell}^{EE}$. The resulting 
pdf has the form
\begin{eqnarray}
f(D_{\ell}^{TT},D_{\ell}^{TE})= n^2{x}^{\frac{n-3}{2}}
\left\{2^{1+n}\pi\Gamma^2(\frac{n}{2})(1-\rho_{\ell}^2)(\sigma_{\ell}^T)^{2n}
(\sigma_{\ell}^E)^2\right\}^{-\frac{1}{2}}
\nonumber\\
 \times\exp\left\{\frac{1}{1-\rho^2_{\ell}}\left(\frac{{\rho_{\ell}}
{z}}{{\sigma_\ell^T}{\sigma_\ell^E}}-\frac{{z}^2}{2x{(\sigma_\ell^E)^2}
}-\frac{{x}}{2{(\sigma_\ell^T)^2}}\right)\right\}.
 \label{pdf_CT}
\end{eqnarray}

In order to estimate parameters, such as $R$, $A_s$, $n_s$, from observations
one seeks the maximum of the likelihood function. The likelihood function is 
the pdf in which the data (estimates $D_{\ell}^{XX'}$) are known while the 
parameters are unknown. Up to a normalization constant, the likelihood 
function is  
\begin{eqnarray}
\nonumber\label{ctlikelihood1}
 \mathcal{L}\propto \prod_{\ell}f(D_{\ell}^{TT}, D_{\ell}^{TE})
 \end{eqnarray}
for $\ell=2,\cdot\cdot\cdot,\ell_{max}$. It can be rewritten as 
 \begin{eqnarray}
 \label{ctlikelihood2}
 -2\ln \mathcal{L}=\sum_{\ell}\left\{\frac{1}{1-\rho^2_{\ell}}
\left(\frac{{z}^2}{x{(\sigma_\ell^E)^2} }+
\frac{{x}}{{(\sigma_\ell^T)^2}} -\frac{{2\rho_{\ell}}{z}}
{{\sigma_\ell^T}{\sigma_\ell^E}}\right) +
\ln\left((1-\rho_{\ell}^{2})(\sigma_{\ell}^T)^{2n}(\sigma_{\ell}^E)^2\right)
\right\}+C,
 \end{eqnarray} 
where the constant $C$ is chosen to make the maximum value of $\mathcal{L}$ 
equal to 1.

Our previous analysis \cite{zbg} was based on the background $\Lambda$CDM 
cosmological model as derived in \cite{wmap5old}. In addition to the relation 
$n_t=n_s-1$, the perturbation parameters $A_s, A_t$ were restricted by 
the observational condition 
$\ell(\ell+1)C_{\ell}^{TT}/2\pi|_{\ell=10}=840~\mu{\rm K}^2$. One more 
restriction was supplied by the phenomenological relation 
$n_s(R)=0.96 +0.35R-0.07R^2$ which indirectly took into account the data on 
$TT$ anisotropies. The remaining free parameter $R$ was subject to the 
likelihood analysis. This analysis was directly using the 5-year WMAP $TE$ 
data at multipoles $2\leq\ell\leq100$ \cite{LAMBDA}. The noise power spectra 
$N_{\ell}^{TT}$, $N_{\ell}^{EE}$ were obtained from the information 
posted at \cite{LAMBDA} and were presented as graphs in \cite{zbg} (Fig.~6). 

The likelihood procedure has resulted in $R=0.240^{+0.291}_{-0.225}$ 
(68.3\% C.L.). For the ML value $R=0.240$, the imposed restrictions have
produced the full set of perturbation parameters:
\begin{equation}
\label{oldML}
R= 0.240,~~n_s=1.040,~~A_s=2.034\times10^{-9},~~A_t=0.960\times10^{-9}
\end{equation}
and $n_t=0.040$. A shift of the parameter $R$ within its confidence 
interval would automatically produce a change in other parameters too. 

In order to avoid any association with inflationary 
predictions, we are not using the parameter $r$. However, if $r$ is 
defined as $r\equiv A_t/A_s$ (definition used by the WMAP team) without 
implying inflationary formulas (\ref{tsr}), then one can establish a 
relation between $R$ and $r$ which depends on the background cosmological 
model and spectral indices \cite{Rossetta}. We derived this relation 
numerically. For a rough comparison of results one can use $r\approx2R$.


\subsection{Revised analysis of the WMAP5 data}

In this paper, the previous approach \cite{zbg} is improved in two main 
aspects. First,
we work directly with both datasets, $TT$ and $TE$. Second, in the 
likelihood procedure all three parameters $R, A_s, n_s$ are kept free. 
(We tried to include $n_t$ as a free parameter, but this did not change 
the results except increasing uncertainties around the ML values.) 

As before, the WMAP5 estimates for $D_{\ell}^{TT}$, $D_{\ell}^{TE}$ at 
multipoles 
$\ell\le\ell_{max}=100$ are taken from \cite{LAMBDA}. The noise power 
spectra $N_{\ell}^{TT}$, $N_{\ell}^{EE}$ are the same as derived 
in \cite{zbg} (Fig. 6). Numerical evaluations of the CMB power spectra 
are performed with the help of CAMB code \cite{CAMB}.
 
The adopted background model is the best-fit $\Lambda$CDM cosmology 
\cite{wmap5} (ApJS version) with parameters 
\begin{eqnarray} 
 \label{background}
 \begin{array}{c} 
 \Omega_bh^2=0.02267^{+0.00058}_{-0.00059},~\Omega_ch^2=0.1131\pm0.0034, \\
 \Omega_{\Lambda}=0.726\pm0.015,~\tau_{\rm reion}=0.084\pm0.016,~
h=0.705\pm0.013.
 \end{array}
\end{eqnarray}
In numerical calculations, we use the central values of these parameters. 

\begin{figure}
\begin{center}
\includegraphics[width=6cm,height=7cm]{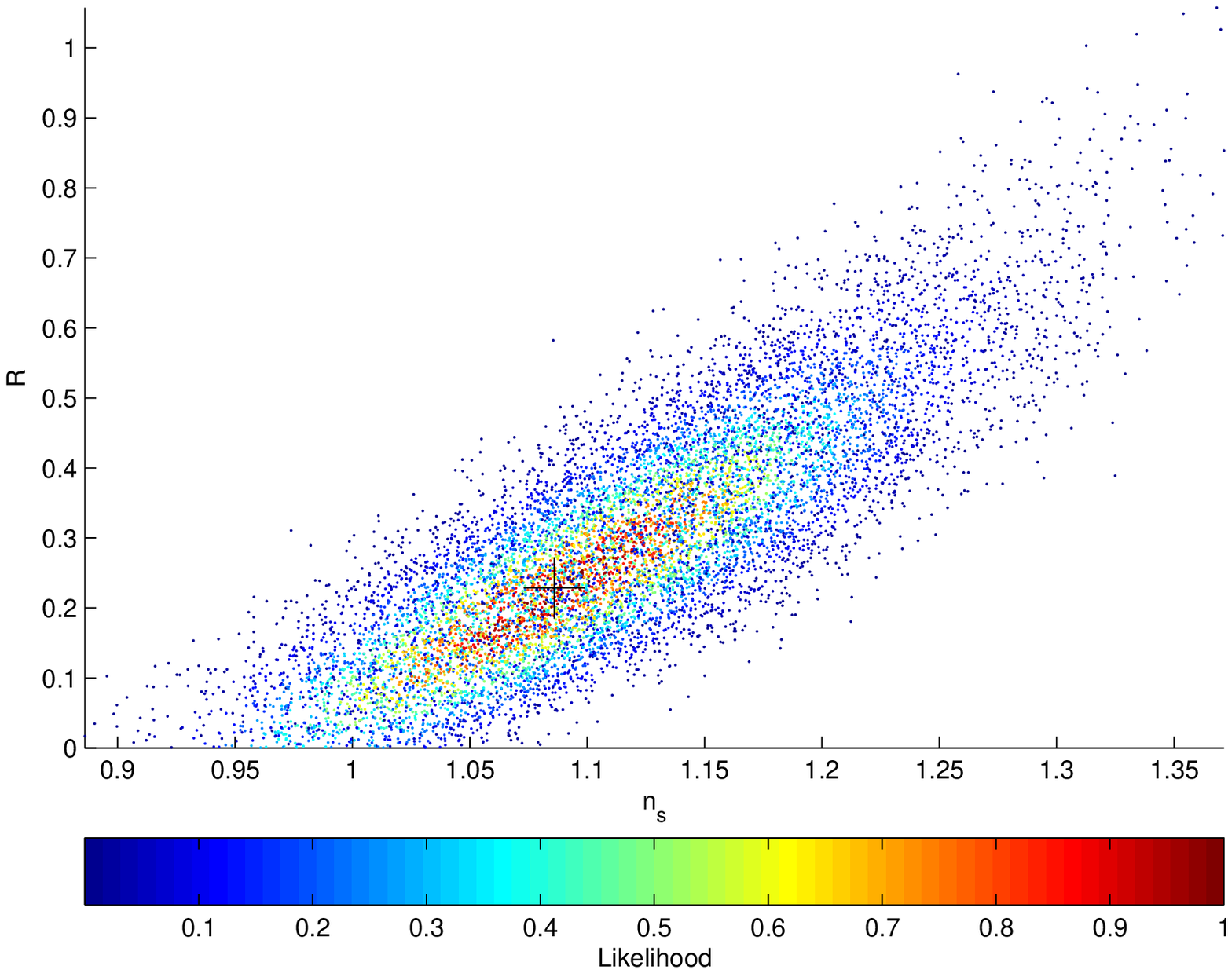}\includegraphics[width=6cm,height=7cm]{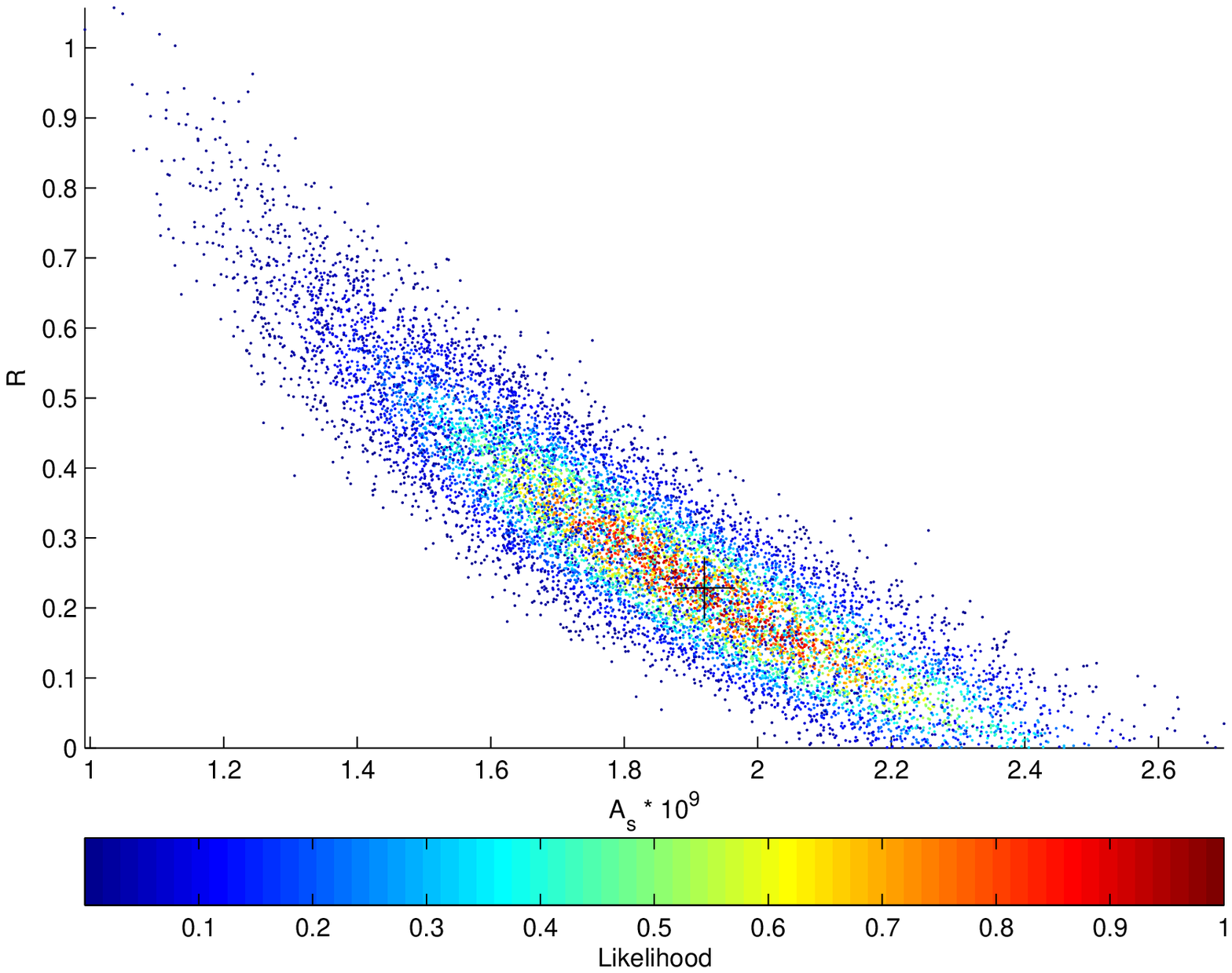}
\end{center}
\caption{The projection of 10,000 samples of the 3-dimensional likelihood 
function onto the  $R-n_s$ (left panel) and $R-A_s$ (right panel) planes. 
The black $+$ indicates the parameters listed in (\ref{best-fit}). }
\label{figurea1.1}
\end{figure}

Applying the Markov Chain Monte Carlo (MCMC) method (see, for example, 
\cite{mcmc1,mcmc2}), we probe the likelihood function (\ref{ctlikelihood2}) 
by 10,000 samples and determine the position of its maximum in 3-dimensional 
space $R, A_s, n_s$. The parameters of our best-fit model, i.e. the maximum 
likelihood (ML) values of the perturbation parameters, are found to be
\begin{eqnarray}
R=0.229,~~~n_s=1.086,~~~A_s=1.920\times10^{-9}
\label{best-fit}
\end{eqnarray} 
and $n_t=0.086$. Obviously, these are only the coordinates of the maximum
in the parameter space. There are many neighbouring points with almost 
equally large values of the likelihood $\mathcal{L}$. It is difficult to 
visualize the 3-dimensional region around the maximum, so in Fig.~\ref{figurea1.1}
we show the projection of the 10,000 sample points on the 2-dimensional planes $R-n_s$ and $R-A_s$.

The color of an individual point  in Fig.~\ref{figurea1.1} 
signifies the value of the 3-dimensional likelihood of 
the corresponding sample. The projections of the maximum (\ref{best-fit}) are
shown by a black $+$. The samples with relatively large values of the 
likelihood (red, yellow and green colors) are concentrated along the 
curve, which projects into relatively straight lines (at least, up to 
$R \approx 0.5$): 
\begin{eqnarray}
\label{1Dmodel}
 n_s=0.98+0.46R, ~~~~~A_s=(2.27-1.53R)\times10^{-9}.
\end{eqnarray}
These combinations of the parameters $R, n_s, A_s$ produce roughly equal 
responses in the CMB power spectra. The best-fit model (\ref{best-fit}) is a 
particular point on these lines, $R=0.229$. We will be using this 
one-parameter family of models (\ref{1Dmodel}) in our study of the detection 
abilities of the {\it Planck} mission in Section \ref{section4}.

Before comparing the new and old results, it is instructive to explore the 
marginalized 2-dimensional and 1-dimensional distributions. The marginalized
distribution over a parameter is the integral of the likelihood function
over that parameter. By integrating $\mathcal{L}$ (\ref{ctlikelihood2}) over
$A_s$ or $n_s$ we derive 2-dimensional likelihoods in $R-n_s$ or $R-A_s$ 
spaces. Then we apply the standard procedure of finding the maxima and 
confidence contours.   

\begin{figure}
\begin{center}  
\includegraphics[width=6cm,height=6cm]{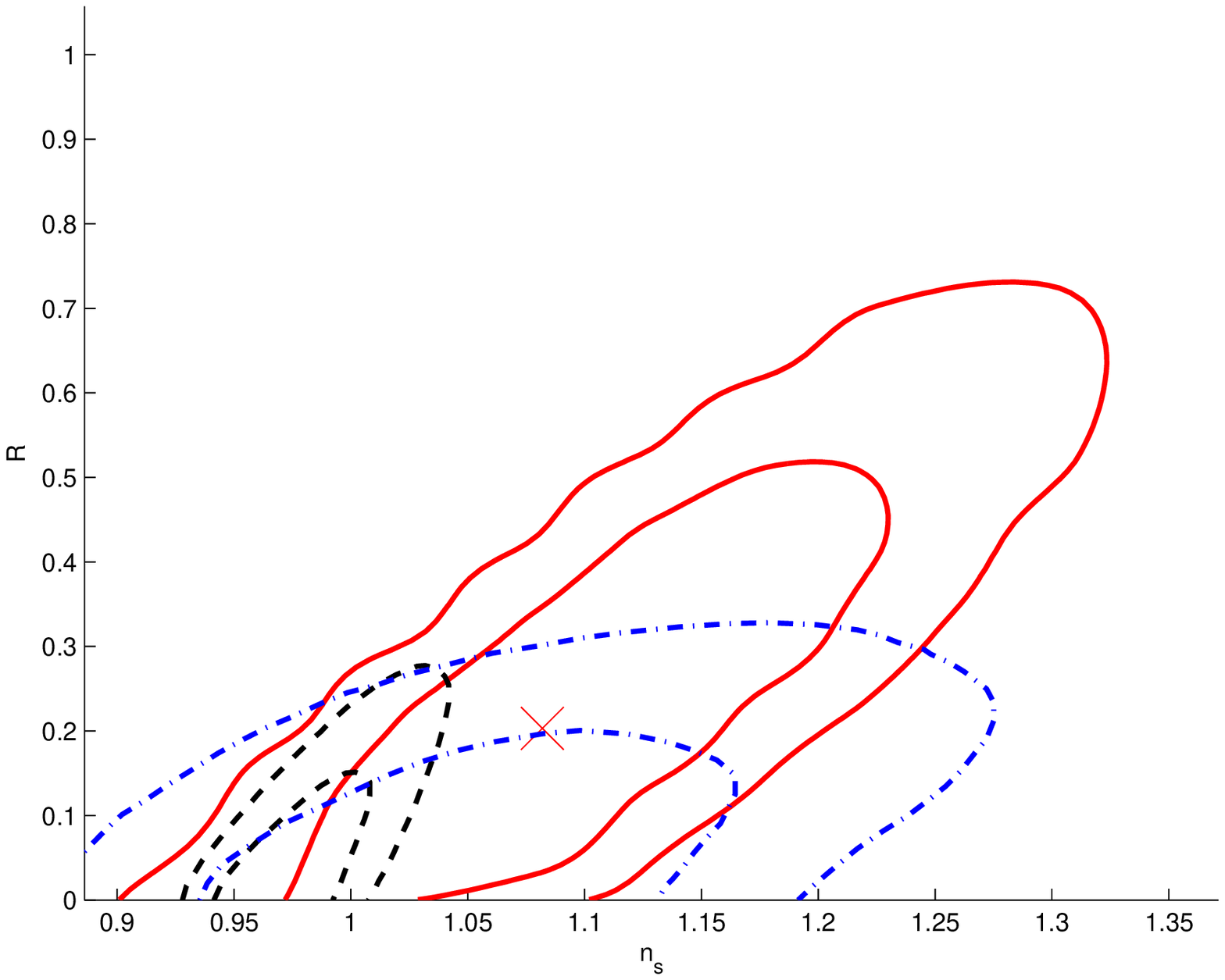}\includegraphics[width=6cm,height=6cm]{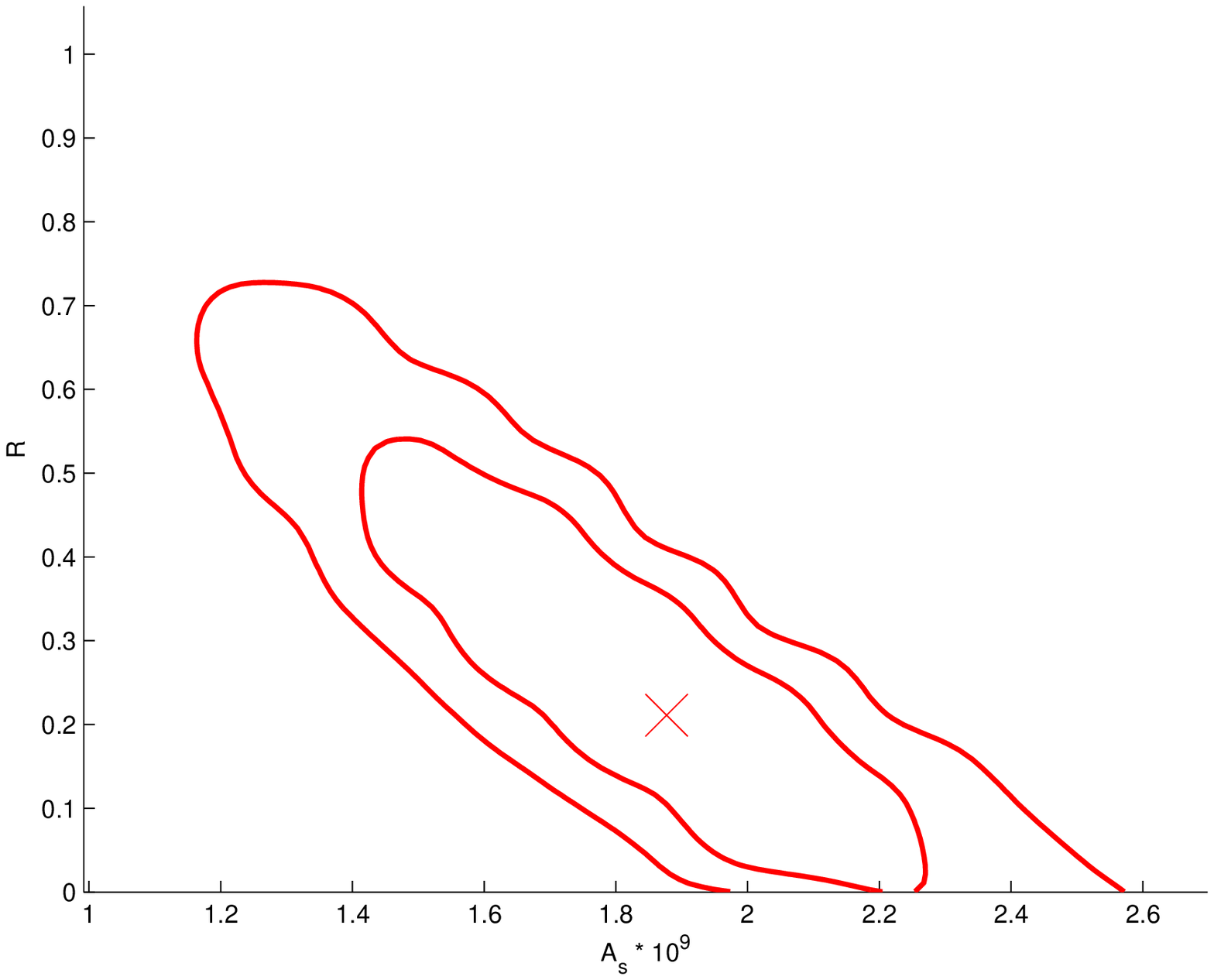}
\end{center}
\caption{The ML points (red $\times$) and the $68.3\%$ and $95.4\%$ confidence 
contours (red solid lines) for 2-dimensional likelihoods: $R-n_s$ (left panel) 
and $R-A_s$ (right panel). In the left panel, the WMAP confidence contours 
are also shown for comparison.}
\label{figurea1}
\end{figure}

In the left panel of Fig.~\ref{figurea1} we show the ML point (marked by
a red $\times$) and the $68.3\%$ and $95.4\%$ confidence contours (red solid 
lines) in the $R-n_s$ plane. The 2-parameter maximum is located at  
 \begin{eqnarray}\label{best-fit-2.1}
  R=0.203,~~n_s=1.082.
 \end{eqnarray}
In the same panel we show 2-dimensional confidence contours as given by the 
WMAP team \cite{wmap5}. We transferred their contours originally plotted in 
$r-n_s$ plane to our $R-n_s$ plane using the numerical relation between $R$ 
and $r$. Their contours are based either on the assumed constancy of the 
spectral index $n_s$ throughout all the explored multipoles (black dashed 
curves), or on a simple running of $n_s$, 
$n_s(k)=n_s(k_0)+\frac{dn_s}{d\ln k}\ln(\frac{k}{k_0})$ 
(blue dash-dotted curves).  

In the right panel of Fig.~\ref{figurea1} we show the ML point (marked by
a red $\times$) and the $68.3\%$ and $95.4\%$ confidence contours (red solid 
lines) in the $R-A_s$ plane. The 2-parameter maximum is located at   
 \begin{eqnarray}\label{best-fit-2.2}
  R=0.211,~~A_s=1.877\times10^{-9}.
 \end{eqnarray}

\begin{figure}
\begin{center}
\includegraphics[width=5cm]{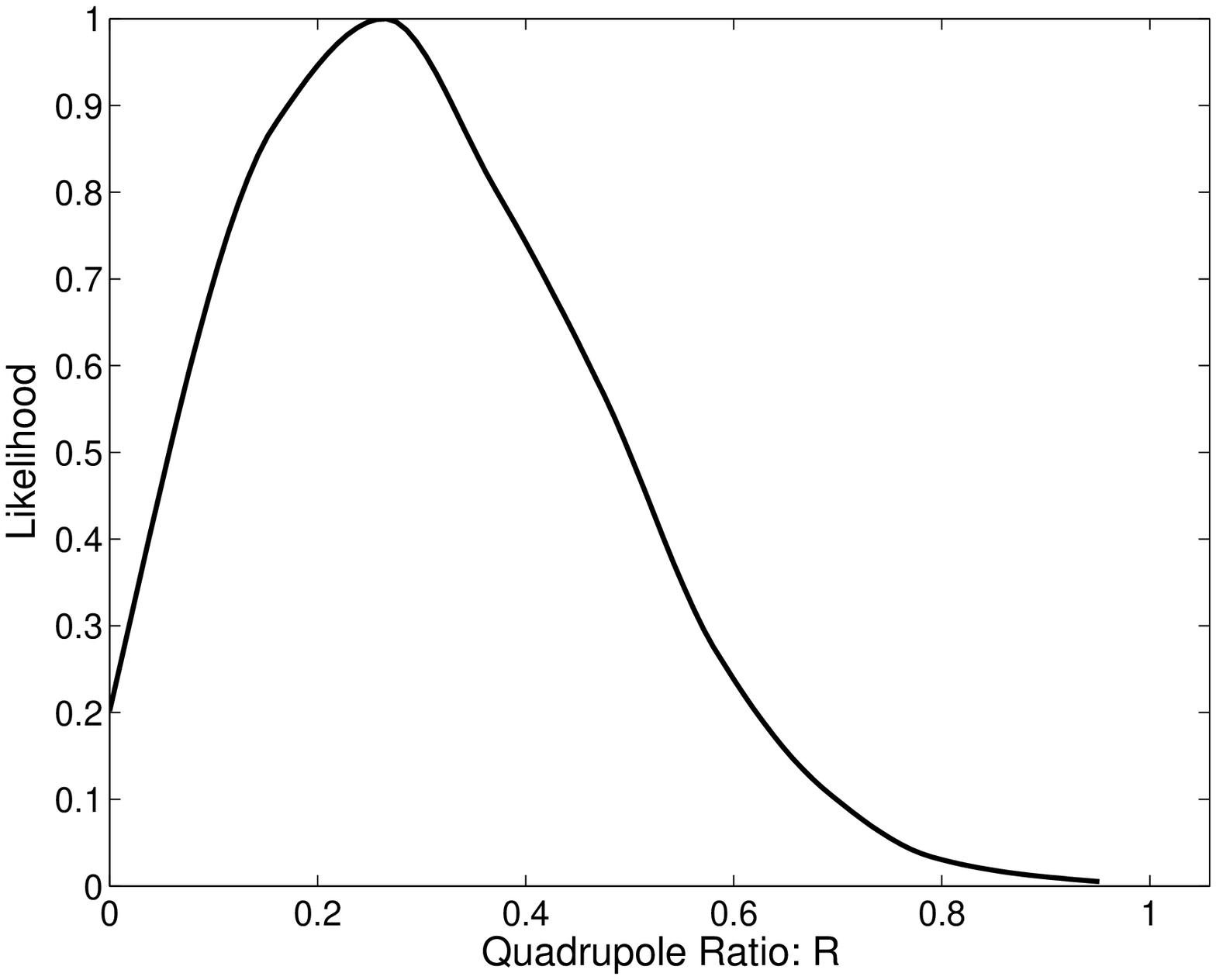}\includegraphics[width=5cm]{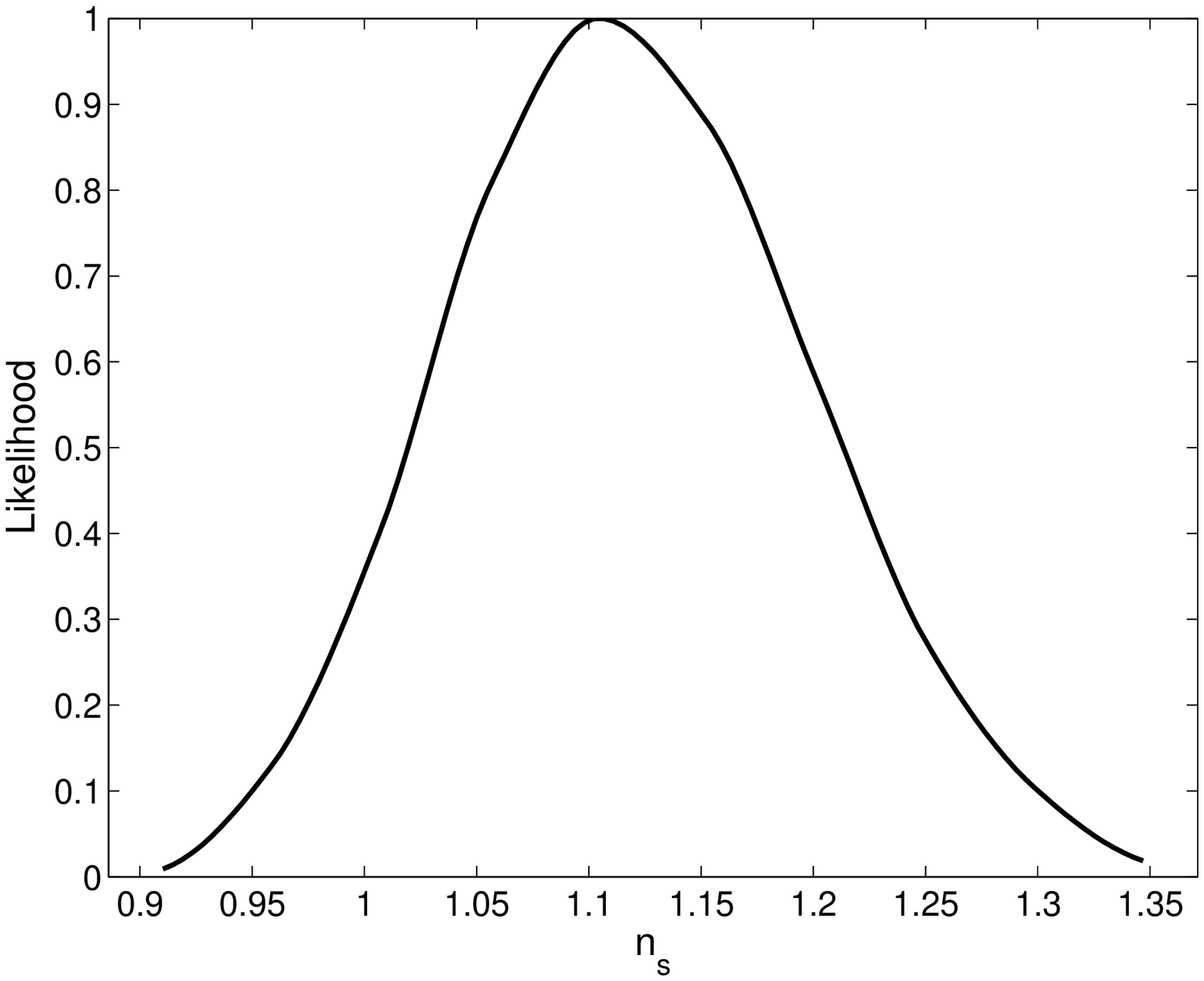}\includegraphics[width=5cm]{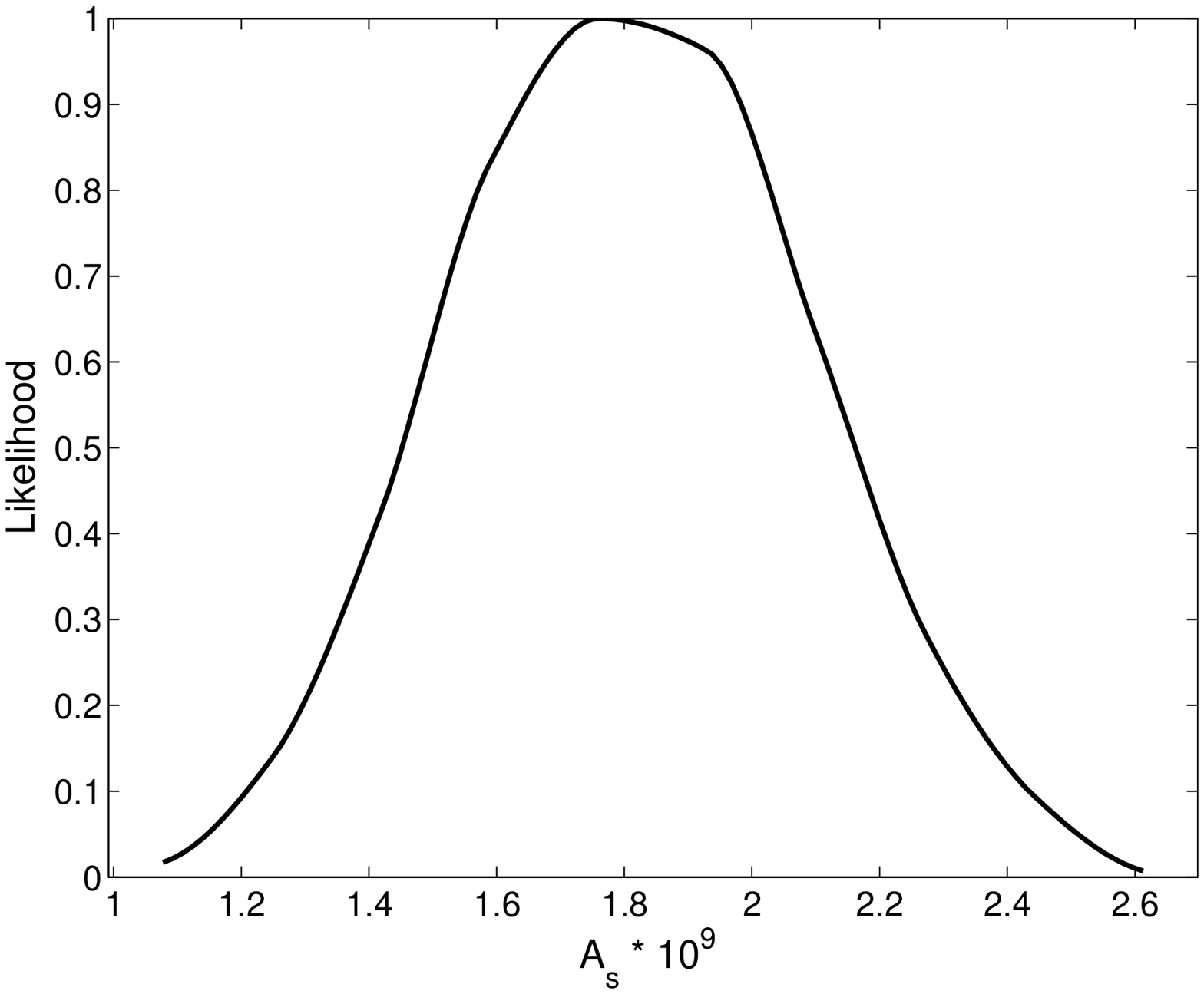}
\end{center}\caption{1-dimensional likelihoods for $R$ (left), $n_s$ (middle) 
and $A_s$ (right).}\label{figureb12}
\end{figure}

Integrating the likelihood function $\mathcal{L}$ (\ref{ctlikelihood2}) over 
two parameters ($A_s$, $n_s$), ($A_s$, $R$) or ($R$, $n_s$), we arrive at 
1-dimensional distributions for $R$, $n_s$ or $A_s$, respectively. We plot
these distributions in Fig.~\ref{figureb12}. The ML values of these parameters
and their $68.3\%$ confidence intervals are given by  
\begin{eqnarray}\label{best-fit-1d}
  R=0.266\pm0.171,~~n_s=1.107^{+0.087}_{-0.070} , 
~~A_s=(1.768^{+0.307}_{-0.245})\times10^{-9}.
\end{eqnarray} 

Comparing the old, Eq.~(\ref{oldML}), and new, Eqs.~(\ref{best-fit}), 
(\ref{best-fit-2.1}), (\ref{best-fit-2.2}), (\ref{best-fit-1d}), results, one 
can conclude the following. First, all the results are close to each
other and deviate little around the rigorous 3-dimensional ML values 
(\ref{best-fit}). Second, the parameter $R$ persistently indicates a 
significant amount of relic gravitational waves, even if with a 
considerable uncertainty. The $R=0$ hypothesis (no gravitational 
waves) appears to be away from the $R=0.229$ model at about $1\sigma$ 
interval, or a little more, but not yet at a significantly 
larger distance. Third, the spectral indices $n_s, n_t$ persistently point 
out to the `blue' shape of the primordial spectra, i.e. $n_s >1, n_t >0$, in 
the interval of wavelengths responsible for the analyzed multipoles    
$\ell\le\ell_{max}=100$. This puts in doubt the (conventional) 
scalar fields as a possible driver for the initial kick, because the scalar 
fields cannot support $\beta > -2$ in Eq.~(\ref{inscf}) and, consequently, 
$n_s >1, n_t >0$ in Eq.~(\ref{PsPt}).


\subsection{Quadrupole data and extrapolation of the ML model to higher 
multipoles}

It is known \cite{quadrupole1} that the actually observed quadrupole 
$D_{\ell=2}^{TT}$ has anomalously low value in comparison with the usually 
plotted graphs of the best-fit CMB power spectra. Since our results prefer 
a somewhat `blue' primordial spectrum, the natural question arises whether 
the low value of the quadrupole is not the reason entirely responsible 
for our evaluation. In order to answer this question we have conducted the 
likelihood analysis without using the observed data points $D_{\ell=2}^{TT}$ 
and $D_{\ell=2}^{TE}$. We found that even this drastic measure of complete 
removal of these data points does not change our results qualitatively. The 
parameters of the maximum likelihood model modify to $R=0.190$, $n_s=1.067$ 
and $A_s=1.993\times10^{-9}$. This is one more evidence of the stability of
indications on the presence of relic gravitational waves in the WMAP5 CMB 
data.

As was already stressed in the paper, we analyze only those WMAP5 data where 
one can expect to find relic gravitational waves, that is, in the range 
$2\leq\ell\leq100$. Therefore, the parameters (\ref{best-fit}) apply only to 
wavelengths responsible for that range. We will show in 
Sec.~\ref{section3} that it can be misleading to try to constrain gravitational
waves by the data outside this interval of multipoles, as the spectral indices 
may change. Nevertheless, it is interesting to see what kind of CMB power 
spectra the ML model (\ref{best-fit}) generates, if the spectral indices are 
assumed fixed at their values (\ref{best-fit}) throughout all the relevant 
wavelengths. 

\begin{figure}
\begin{center}
\includegraphics[width=16cm,height=9cm]{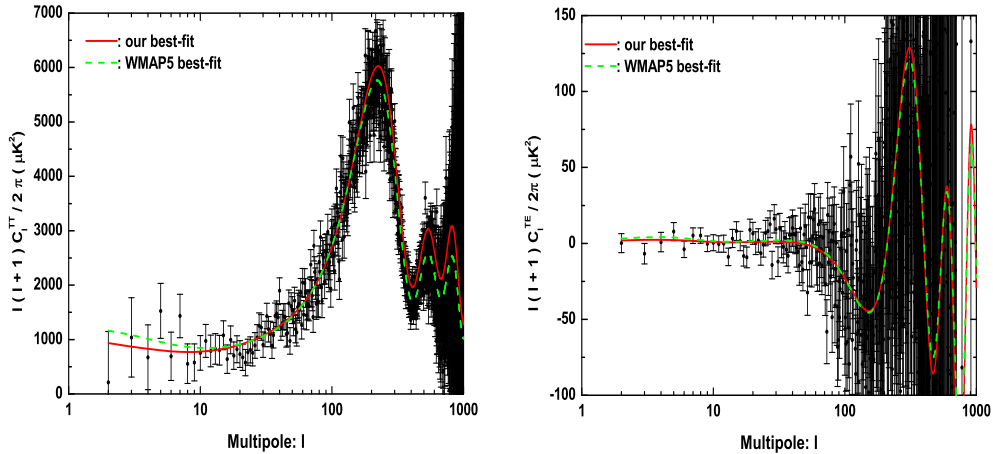}
\end{center}\caption{The unbinned WMAP5 $TT$ (left panel) and $TE$ (right 
panel) data \cite{LAMBDA}, along with their uncertainties due to noises and 
statistics. The red solid lines show the ML model (\ref{best-fit}) 
extrapolated with constant spectral indices to high multipoles. For 
comparison, the dashed green line shows the WMAP5 best-fit ``no gravitational 
waves" model.}
\label{figurec1}
\end{figure}

In  Fig.~\ref{figurec1}, we show these extrapolated $TT$ and $TE$ power 
spectra built on the ML parameters (\ref{best-fit}). One can see that these 
spectra are not too far away from the ``no gravitational waves" spectra 
advocated by the WMAP team. As one could expect, the somewhat `blue' spectral 
index $n_s$ in (\ref{best-fit}) makes the extrapolated spectra 
positioned somewhat above the WMAP5 spectra at very large multipoles.

\section{How relic gravitational waves can be overlooked in the likelihood 
analysis of $TT$ and $TE$ data\label{section3}}

With all the reservations already stated, our results are markedly different 
from the WMAP5 conclusions \cite{wmap5}. The WMAP team has found no evidence 
for gravitational waves and arrived at a red spectral index $n_s=0.96$. The 
WMAP findings are symbolized by black dashed and blue dash-dotted contours
in Fig.~\ref{figurea1}. It is important to understand the reasons for these 
disagreements.  

Two differences in data analysis have already been mentioned. We restrict
our analysis to multipoles $\ell\leq100$, whereas the WMAP team uses 
the data at all multipoles up to $\ell\sim1000$ keeping spectral indices 
constant. We use the relation $n_t=n_s-1$ implied by the theory
of quantum-mechanical generation of cosmological perturbations, whereas 
the WMAP team uses the inflationary `consistency relation' $r=-8n_t$
which automatically sends $r$ to zero when $n_t$ approaches zero. There
could be some discrepencies in treating the noises, but we think
we effectively followed \cite{zbg} the WMAP prescriptions. After several
trials, we came to the conclusion (with heavy heart, as Einstein used to 
say) that it is the assumed constancy of spectral indices in a broad
spectrum that is mostly responsible for the disagreement, and it should 
be abandoned. The constancy of $n_s$ over the vast region of wavenumbers,
or possibly a simple running of $n_s$, is a usual assumption in a number
of works \cite{wmap5}, \cite{othergroups,wishart3}. 

\begin{figure}
{\includegraphics[width=6cm,height=6cm]{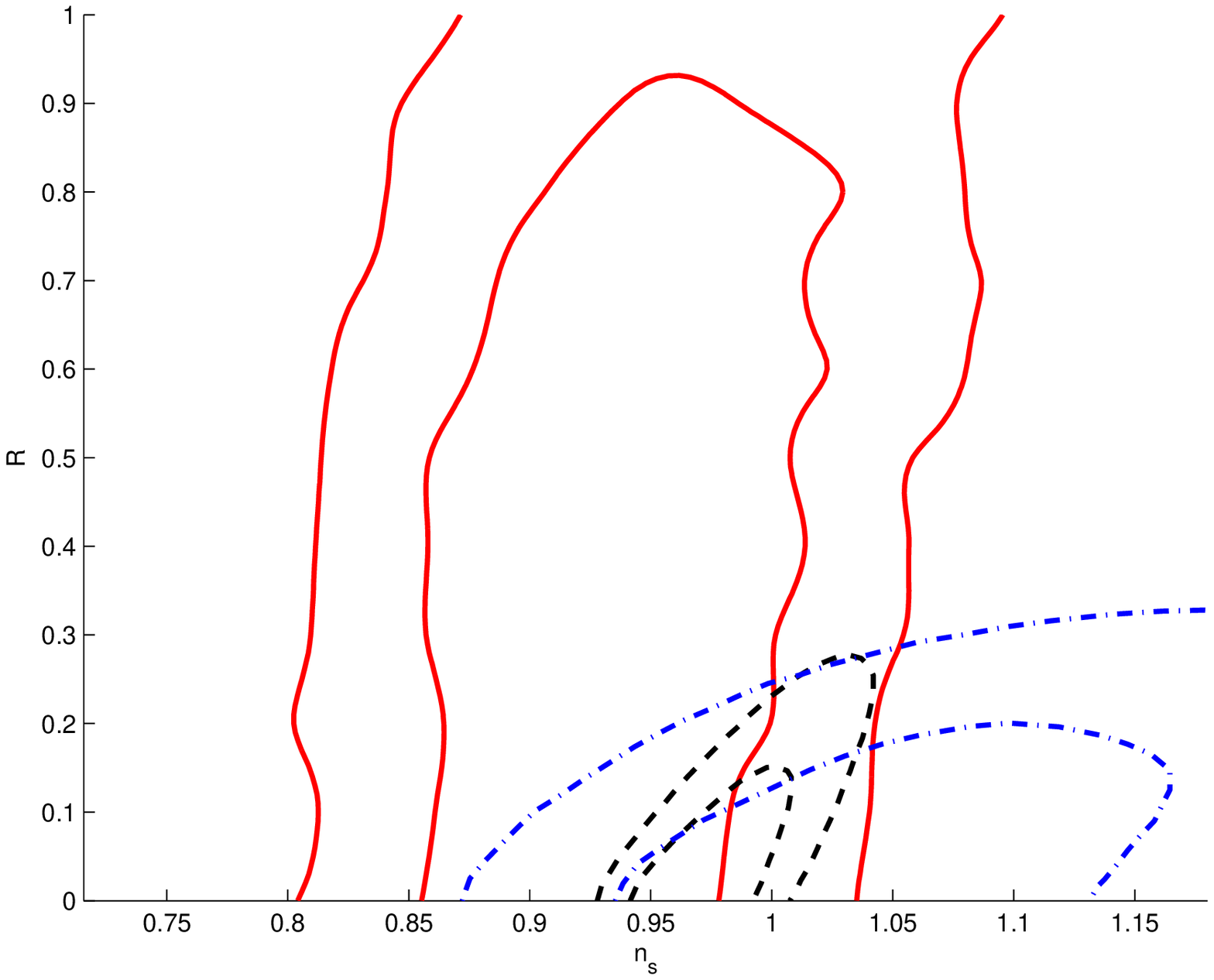}}
{\includegraphics[width=6cm,height=6cm]{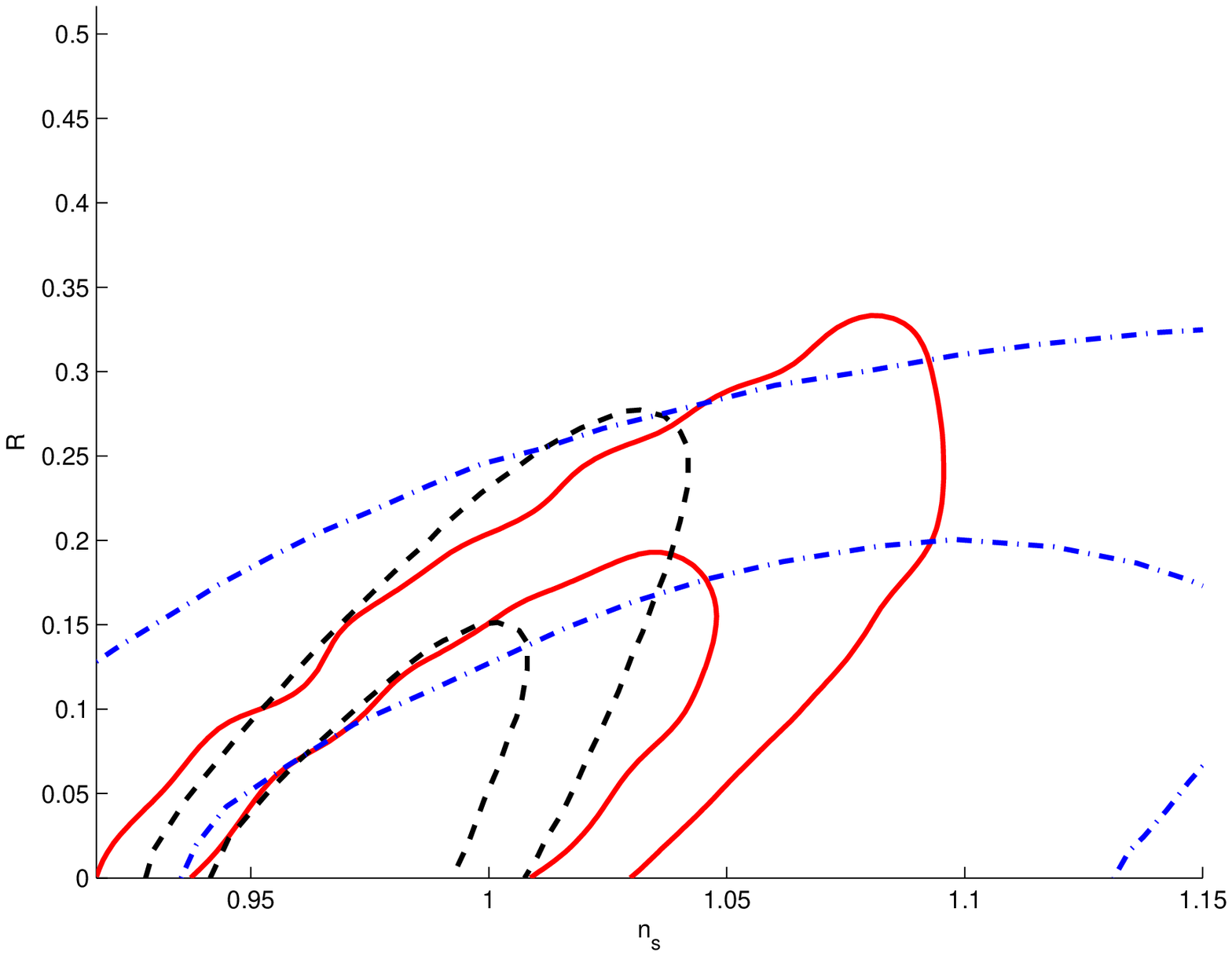}}
\caption{Same evaluations as in the left panel in Fig.~\ref{figurea1}, but 
derived from WMAP5 $TT+TE$ data in the interval $101\le\ell\le220$ 
(left panel), and in the interval $2\le\ell\le220$ (right panel).
WMAP5 contours \cite{wmap5} are shown for comparison.} \label{sec3-fig1}
\end{figure}

\begin{figure}
{\includegraphics[width=6cm,height=6cm]{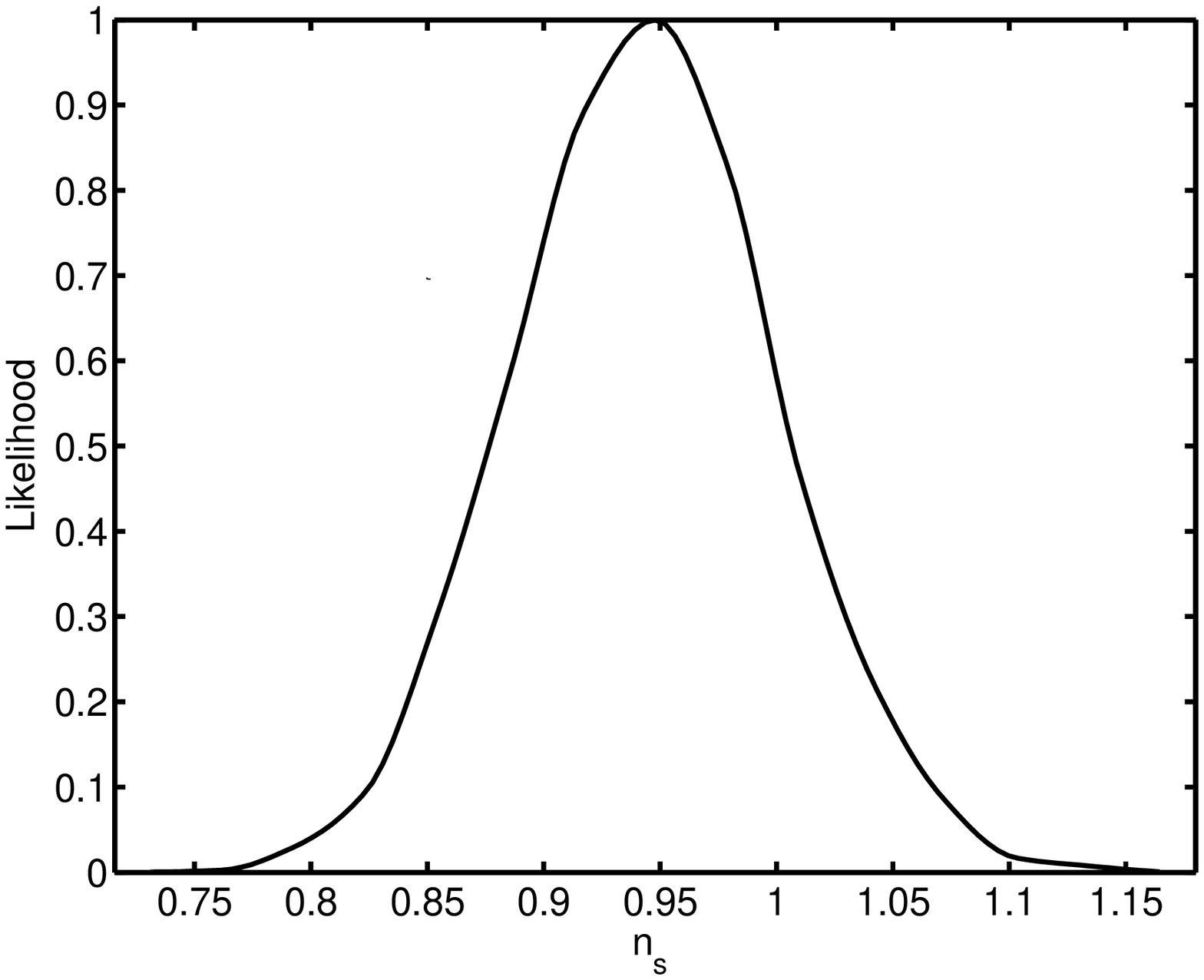}}
{\includegraphics[width=6cm,height=6cm]{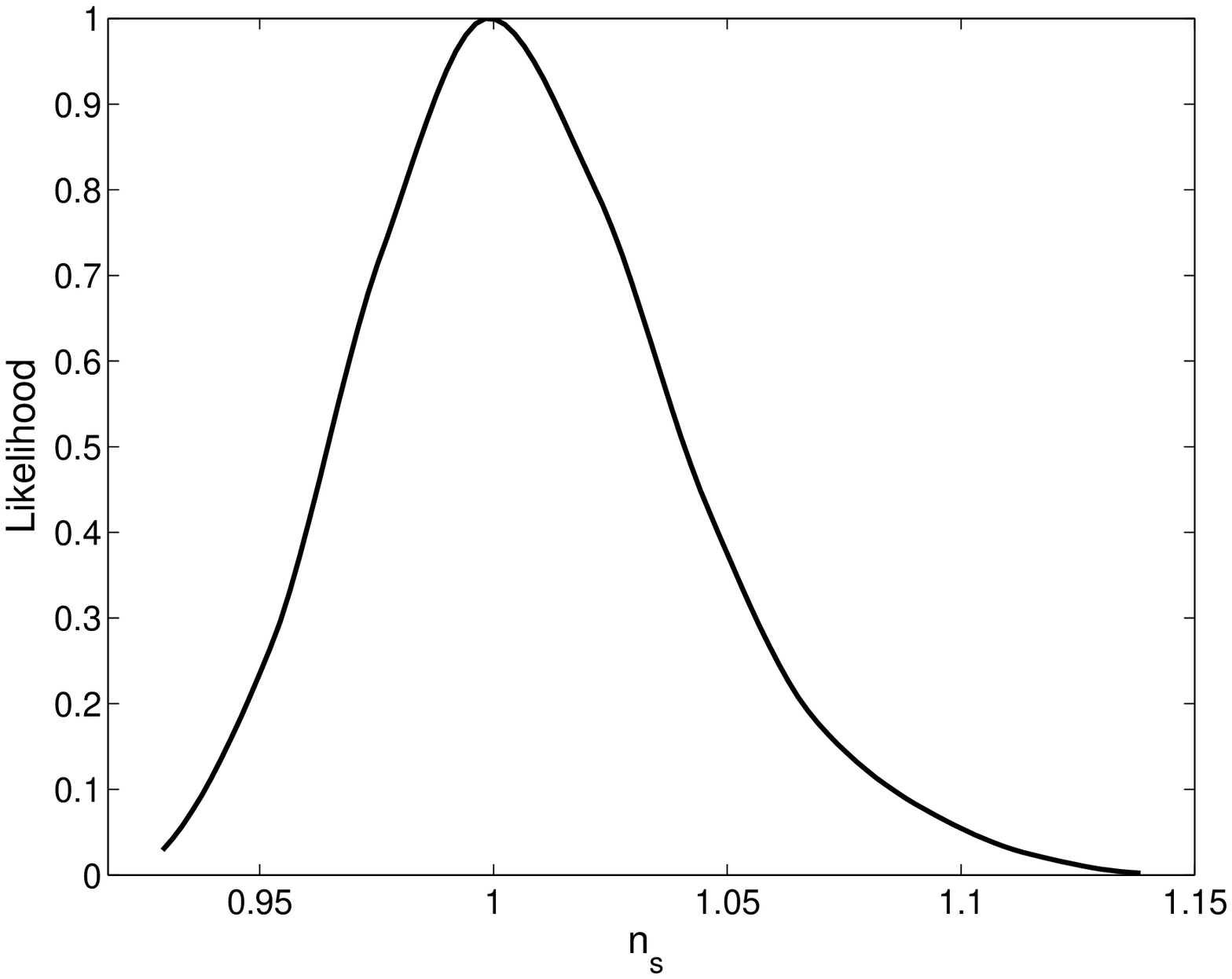}}
\caption{1-dimensional likelihood for $n_s$ derived from WMAP5 $TT+TE$ data
in the interval $101\le\ell\le220$ (left panel), and
in the interval $2\le\ell\le220$ (right panel).}\label{sec3-fig2}
\end{figure}

In order to understand the impact of higher multipole data, 
we first probed the likelihood function and estimated the parameters from 
the data in the range of multipoles $101\le\ell\le220$. The procedure was
exactly the same as was used in the analysis of $2\le\ell\le100$ data.
The maximum of the 3-parameter likelihood function was found at 
$n_s=0.923$, $R=0.022$ and $A_s=2.65\times10^{-9}$, that is, at a
distinctly `red' spectral index $n_s$. The 2-dimensional marginalized
distribution $R-n_s$ (analogous to the left panel in Fig.~\ref{figurea1}) 
is also different. It is shown in the left panel of Fig.~\ref{sec3-fig1}. 
The large uncertainty surrounding $R=0.022$ reflects the fact that, at 
these multipoles, the contribution of relic gravitational waves becomes 
very small. It is the density perturbations that play dominant role here,
and at higher multipoles. The 1-dimensional marginalized distribution for 
$n_s$ (analogous to the middle panel in Fig.~\ref{figureb12}) is shown in the 
left panel of Fig.~\ref{sec3-fig2}. This distribution gives 
$n_s=0.948^{+0.052}_{-0.061}$ ($68.3\%$ C.L.). Obviously, this value of  
$n_s$ is significantly smaller than the one in Eq.~(\ref{best-fit-1d}), 
and the two evaluations do not overlap in 1$\sigma$ confidence interval. 
This is a clear indication that the spectral index $n_s$ can hardly be 
treated as one and the same constant throughout all the wavelengths 
responsible for $2\le\ell\le100$ and $101\le\ell\le220$ intervals.

As one could expect, exactly the same analysis of the whole interval 
$2\leq\ell\leq220$ from the position of constant $n_s$ leads to the 
ML values which are intermediate between evaluations at the intervals
$2\le\ell\le100$ and $101\le\ell\le220$ separately. The 3-parameter 
likelihood analysis applied to WMAP5 $TT+TE$ data in the interval
$2\leq\ell\leq220$ has resulted in the ML values $n_s=0.973$, $R=0.019$ 
and $A_s=2.39\times10^{-9}$. As expected, the ML value $n_s=0.973$ is in 
between the ML values $n_s=1.086$ and $n_s=0.923$ from the two adjacent
intervals of $\ell$. The 2-dimensional marginalized constraints are 
shown in the right panel of Fig.~\ref{sec3-fig1}. In comparison with the
left panel, the uncertainty contours are much closer to the WMAP5 evaluations. 
The 1-dimensional marginalized distribution for $n_s$ is shown in the right 
panel of Fig.~\ref{sec3-fig2}. It gives $n_s=0.998^{+0.036}_{-0.029}$ 
($68.3\%$ C.L. ). Again, this is the intermediate value of $n_s$ in 
comparison with evaluations from the two intervals of $\ell$ separately.
Our evaluation of $n_s$ from the data in the interval $2\le\ell\le220$
still gives a little bit higher value of $n_s$ than $n_s=0.96$ found by the 
WMAP team, but presumably the remaining difference is accounted for by the 
multipoles $\ell > 220$ and other data sets that were included in the WMAP derivation. 

One can see now why the inclusion of data from $\ell > 100$ is 
dangerous. Although these data have nothing to do with gravitational 
waves, they bring $n_s$ down. If one assumes that $n_s$ is one and the 
same constant at all $\ell$'s, this additiosnal `external' information 
about $n_s$ affects uncertainty about $R$ and brings $R$ down. This is 
clearly seen, for example, in the left panel of Fig.~\ref{figurea1}. 
The localization of $n_s$ near, say, the line $n_s =0.96$ would cross 
the solid red contours along that line and would enforce lower, or 
zero, values of $R$. However, as we have shown above, the $n_s$ is 
sufficiently different even at the span of the two neighbouring intervals 
of $\ell$'s that we discussed. These considerations as for how relic
gravitational waves can be overlooked have general significance and
will be applicable to any CMB data and data analysis.


\section{Forecasts for the {\it Planck} mission \label{section4}}

Now, that {\it Planck} satellite \cite{planck} is launched, it is 
important to see in detail how the maximum likelihood model (\ref{best-fit}) 
found from WMAP5 data, as well as neighbouring models defined by 
Eq.~(\ref{1Dmodel}), will fare in the less noisy {\it Planck} data. 
In our previous work \cite{zbg}, we restricted our consideration to only two 
separate information channels, namely to $TE$ and $BB$ correlation functions.
We took into account only one frequency channel $143\rm{GHz}$ and its 
instrumental noise. We replaced the unknown level of residual foregrounds and 
systematics by the increased total noise in the $BB$ channel, calling it a
`realistic' $BB$ channel. On the data analysis side, we performed the
likelihood analysis with a single parameter $R$, assuming that other
parameters $A_s$, $n_s$ and $n_t = n_s -1$ are known, as long as they are 
related to $R$ by the imposed restrictions.   

In this paper, we study the detection abilities of {\it Planck} in a much more 
comprehensive manner. First, we consider all available information channels, 
i.e.  $TT$, $TE$, $EE$ and $BB$ correlation functions, and their combinations. 
Second, in evaluating the instrumental noises, we take into account all the 
{\it Planck's} frequency channels (either three, or even all seven). Third, 
we include in the total noise the residual foreground contamination from the 
synchrotron and dust emissions, and introduce the `pessimistic' case when the 
foregrounds are not removed at all while the nominal instrumental noise in
the $BB$ channel is increased by a factor of 4. Finally, we search not only 
for a single parameter $R$, but where the computational resources allow, we 
evaluate the uncertainties for $R$ arising in the procedure of joint 
determination of all three unknown parameters $R$, $n_s$, $A_s$ from a given 
set of observational data.


\subsection{The signal-to-noise ratio in the measurement of $R$
\label{section4.1}}

\subsubsection{The noise power spectra and the definition of $S/N$ 
\label{subsection4.1.1}}

Surely, in the focus of our attention is the detection of relic gravitational
waves. Their CMB contribution is parameterized by $R$. To quantify the 
detection ability of a CMB experiment, we introduce the signal-to-noise 
ratio \cite{zbg}
\begin{eqnarray}
\label{snr}
 S/N\equiv R/\Delta R,
\end{eqnarray}
where the numerator is the true value of the parameter $R$ (or its ML value, 
or the input value in a numerical simulation) while $\Delta R$ in the
denominator is the expected uncertainty in determination of $R$ from the 
data. 

In formulating the observational forecasts, it is common to use the Fisher
matrix formalism. We outline its main features in Appendix \ref{appa}. 
The uncertainty $\Delta R$, for a given $R$, depends on noises, statistics 
of the searched for CMB signal (which is random in itself), and the number of 
parameters subject to evaluation from a given dataset. The instrumental and 
foreground noises comprise the total effective noise power spectra 
$N_{\ell}^{XX'}$ which enter the Fisher matrix (\ref{fishermatrix}) and its 
element $F_{RR}$, Eq.~(\ref{FRR}), as well as the element 
${\left(F^{-1}\right)_{RR}}$ of the inverse matrix. Depending on the number 
of sought after parameters, one calculates $\Delta R$ either according to 
Eq.~(\ref{deltaR1}) or according to Eq.~(\ref{deltaR2}). 

The noise power spectra $N_{\ell}^{XX'}$ are explained in Appendix \ref{appb}. 
We ignore the cross-correlated noises, i.e. when $X \neq X'$, so we 
are working only with $N_{\ell}^{XX}$. The instrumental characteristics of
all seven frequency channels \cite{planck} (we mark them by symbol 
$i = 1,\cdot\cdot\cdot,7$) are listed in Table \ref{table1} of Appendix \ref{appb}, 
where LFI means Low Frequency Instrument and HFI High Frequency Instrument.
Following \cite{planck} we treat only three frequency channels 
$100$GHz, $143$GHz and $217$GHz as providing CMB data, whereas other 
channels are supposed to be used for determining the foregrounds. However, in
Section \ref{section4.3} we study the improvements that would arise if one 
could use all seven frequency channels for CMB analysis. It is seen from
(\ref{totalnoise}), that three channels are better than one, and seven 
channels are better than three.

The residual foreground noise $N_{{\rm fg},\ell}^{XX}(i)$ adds to the
instrumental noise $N_{{\rm ins},\ell}^{XX}(i)$ and increases the total noise 
$N_{\ell}^{XX}$ (\ref{totalnoise}). We neglect the effect of foregrounds 
in $TT$ channel because this noise is small in comparison with the
signal \cite{bennett}. The foreground contamination cannot be neglected in
polarization channels. The synchrotron and dust emissions are expected 
to be the dominant hindrances in the {\it Planck} frequency range 
\cite{foreground3,foreground2,cmbpol-fore}. The charecteristics of the
accepted foreground models are listed in Appendix \ref{appb}. In what follows, 
we are using the more severe Dust A model, whereas the more favorable Dust B 
model is considered only in Section \ref{section4.3}. 

The ways of mitigating the foreground contamination are discussed in a number
of papers \cite{remove1,remove2,remove3,remove4,remove5,remove6}. In this 
paper, we take a phenomenological approach and quantify the residual noise 
by the factor $\sigma^{\rm fg}$ (see \cite{foreground3} and 
Appendix \ref{appb}) which multiplies the model power spectra 
$C_{S,\ell}^{XX}(i)$, $C_{D,\ell}^{XX}(i)$ of the synchrotron (S) and 
dust (D) emissions. We consider three cases, $\sigma^{\rm fg}=1$ (no 
foreground removal), $\sigma^{\rm fg}=0.1$ ($10\%$ foreground residual noise) 
and $\sigma^{\rm fg}=0.01$ ($1\%$ foreground residual noise). In order to 
gauge the worst case scenario, we consider also the `pessimistic' case, where
$\sigma^{\rm fg}=1$ and $N_{{\rm ins},\ell}^{BB}(i)$ at each frequency $\nu_i$ 
is 4 times larger than the values listed in Table \ref{table1}. This increased 
noise is meant to mimic the situation where it is not possible to get rid 
of various systematic effects \cite{systematics}, the $E$-$B$ mixture 
\cite{ebmixture}, cosmic lensing \cite{lensing}, etc. which all affect the 
$BB$ channel.

\begin{figure}
\centerline{\includegraphics[width=14cm,height=10cm]{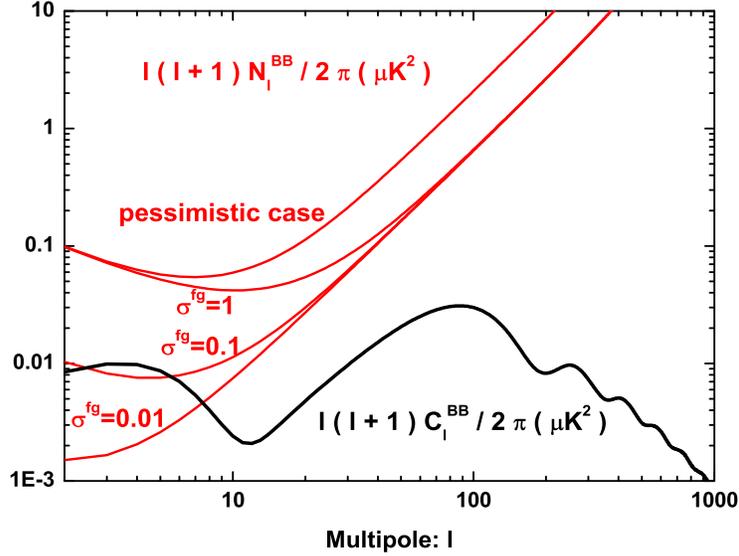}}
\caption{The total noise power spectrum $\ell(\ell+1)N_{\ell}^{BB}/2\pi$
in comparison with the $BB$ power spectrum $\ell(\ell+1)C_{\ell}^{BB}/2\pi $ 
for the model (\ref{best-fit}) with $R=0.229$.}\label{figurev0}
\end{figure}

To illustrate the expected total noise, including the different levels of 
foreground contamination, we show in Fig.~\ref{figurev0} the total noise 
power spectrum $N_{\ell}^{BB}$ calculated according to Eq.(\ref{totalnoise}).
For reference, the black curve shows the power spectrum $C_{\ell}^{BB}$ for 
the maximum likelihood model with $R=0.229$ (see Eq.~(\ref{best-fit})) 
extrapolated up to $\ell =1000$.  It is seen from the graphs that the 
role of foregrounds is restricted to relatively small multipoles 
$\ell\lesssim20$. For higher multipoles, the total noise $N_{\ell}^{BB}$
is dominated by the instrumental noise and does not depend on 
$\sigma^{\rm fg}$. It can also be seen from Fig.~\ref{figurev0} that it
is only for small values of $\sigma^{\rm fg}$, i.e. in the case when the 
foreground contamination is significantly suppressed, that the signal 
$C_{\ell}^{BB}$ will be greater than the noise $N_{\ell}^{BB}$ at 
lowest multipoles. Thus, even in the case of small $\sigma^{\rm fg}$, 
the {\it Planck's} $BB$ channel will be mostly sensitive to gravitational 
waves in the epoch of reionization. For larger values of $\sigma^{\rm fg}$, 
the relative contribution of lower multipoles to the total $S/N$ is 
diminished, signifying the overall reduction of $S/N$. In this case, 
the main contribution to $S/N$ comes from higher multipoles 
$\ell\gtrsim20$ and $S/N$ will ultimately be limited by the level of 
instrumental noise.

\subsubsection{The analysis of the family of models (\ref{1Dmodel}), 
including the $ML$ model (\ref{best-fit})\label{section4.1.2}}

The set of maximum likelihood parameters (\ref{best-fit}) is the best 
set among the `almost equally good' sets, defined by (\ref{1Dmodel}). In a 
sense, Eq.~(\ref{1Dmodel}) is our theoretical prediction, based on the 
analysis of WMAP5 data, of the best viable perturbation models. This family 
of models is the subject of the Fisher matrix analysis below. With all the 
noises $N_{\ell}^{XX}$ defined by (\ref{totalnoise}) and all the power 
spectra $C_{\ell}^{XX'}$ calculable from the family parameters 
(\ref{1Dmodel}), we have enough ammunition to find the quantities 
(\ref{fishermatrix}).   

\begin{figure}
\centerline{\includegraphics[width=18cm,height=10cm]{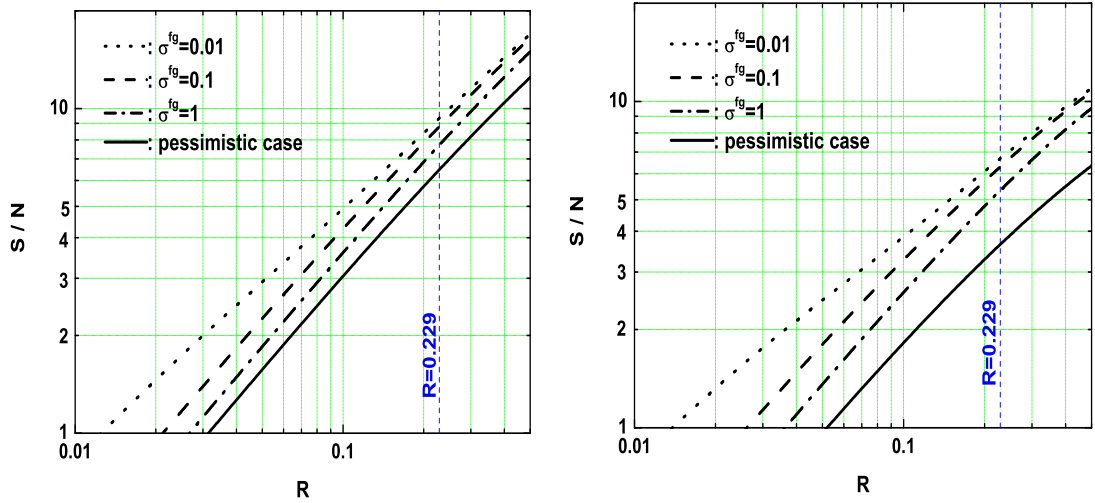}}
\caption{The $S/N$ as a function of $R$. In the left panel, the uncertainty 
$\Delta R$ is calculated according to Eq.~(\ref{deltaR1}), whereas in the
right panel it is calculated according to Eq.~(\ref{deltaR2}).}
\label{figurev2}
\end{figure}

We start from the idealized situation, where only one parameter $R$ is 
considered unknown and therefore the uncertainty $\Delta R$ can be found 
from (\ref{deltaR1}). All the information channels, $TT$, $TE$, $EE$, $BB$, 
are used in the calculation of $F_{RR}$, Eq.~(\ref{FRR}). The results
for $S/N$ are plotted in the left panel of Fig.~\ref{figurev2}. Four options
are depicted, $\sigma^{\rm}=0.01,0.1,1$ and the pessimistic case. 

The results for the benchmark model (\ref{best-fit}) are given by the 
intersection points along the vertical line $R=0.229$. The signal to noise
is high, $S/N=9.35$, if the foregrounds can be removed to the level 
$\sigma^{\rm fg}=0.01$, and $S/N$ decreases to $S/N=8.80$, $S/N=7.72$, 
$S/N=6.48$ for $\sigma^{\rm fg}=0.1$, $\sigma^{\rm fg}=1$, and the pessimistic 
case, respectively.  In all these cases, the $S/N$ is impressively large. 
A value of $R$ down to $R=0.03$ can be measured with $S/N=2$ in the optimistic 
scenario $\sigma^{\rm fg}=0.01$, whereas $S/N>2$ is achieved for 
$R>0.064$ in the pessimistic case. Typically, the optimistic scenario gives 
$S/N$ about 1.5 times larger than the pessimistic case, with the disparity 
growing larger for smaller values of $R$.

As one could expect, the uncertainty $\Delta R$ grows and $S/N$ decreases in 
the realistic situation, when all unknown parameters $R$, $n_s$ and $A_s$ are 
supposed to be evaluated from one and the same dataset. In this case, 
$\Delta R$ should be calculated according to (\ref{deltaR2}). Again, 
calculating $\left(F^{-1}\right)_{RR}$, we take into account all the 
information channels $TT$, $TE$, $EE$, $BB$. The results for $S/N$ are 
presented in the right panel of Fig.~\ref{figurev2}.  For the benchmark model 
(\ref{best-fit}), the signal-to-noise ratios are smaller than in the
left panel: $S/N = 6.69,~6.20,~5.15$ if $\sigma^{\rm fg}=0.01,~0.1,~1$, 
respectively. 

However, the good news is that even in the pessimistic case one 
gets $S/N>2$ for $R>0.11$, and the {\it Planck} satellite will be capable of 
seeing the ML signal $R=0.229$ at the level $S/N=3.65$.

\subsection{Contributions of individual information channels and individual 
multipoles to the total $S/N$}

The evaluations in Sec.~\ref{section4.1.2} assume that all the correlation
functions, $TT+TE+EE+BB$, are taken into account and all the relevant 
multipoles $\ell$ participate in the summations. The total $S/N$ can only 
be worse if something is missing either in the information channels or in the 
accessible multipoles. In order to gain further insight into the detection 
ability of {\it Planck}, it is instructive to make a break-up of $S/N$ over 
combinations of information channels and multipoles. It is unclear how to do 
this in the case of Eq.~(\ref{deltaR2}), but it is relatively easy to do this 
in the case of Eq.~(\ref{deltaR1}). We will limit ourselves to this latter 
case which is sufficient for the purpose of illustration. 

The $S/N$ from Eq.~(\ref{snr}), together with $\Delta R$ from 
Eq.~(\ref{deltaR1}), can be rewritten as  
\begin{eqnarray}\label{s1}
(S/N)^2=R^2 F_{RR},
\end{eqnarray}
The $F_{RR}$ from Eq.~(\ref{FRR}) is a simple sum over multipoles $\ell$, 
so the $(S/N)^2$ in (\ref{s1}) can be decomposed into individul 
$\ell$-contributions
 \begin{eqnarray}\label{s2}
 (S/N)^2=\sum_{\ell}(S/N)_{\ell}^2.
 \end{eqnarray}
Each individul $(S/N)_{\ell}^2$ depends on all information channels, with the 
$BB$-channel factored out, as was infixed from the very beginning in the form
of Eq.~(\ref{ctelikelihood2}),  
\begin{eqnarray}\label{s3}
(S/N)_{\ell}^2&=&\frac{R^2(2\ell+1)f_{\rm sky}}{2\left({\cal C}_{\ell}^{TT}
{\cal C}_{\ell}^{EE}-({\cal C}_{\ell}^{TE})^2\right)^2}\times\left\{({\cal C}_{\ell}^{TT})^2
\frac{\partial {\cal C}_{\ell}^{EE}}{\partial R}\frac{\partial {\cal C}_{\ell}^{EE}}
{\partial R} +({\cal C}_{\ell}^{EE})^2\frac{\partial {\cal C}_{\ell}^{TT}}
{\partial R}\frac{\partial {\cal C}_{\ell}^{TT}}{\partial R}\right.\nonumber\\
 &~&
 +2\left({\cal C}_{\ell}^{TT}{\cal C}_{\ell}^{EE}+\left({\cal C}_{\ell}^{TE}\right)^2\right)
 \frac{\partial {\cal C}_{\ell}^{TE}}{\partial R}\frac{\partial {\cal C}_{\ell}^{TE}}{\partial R} 
+2\left({\cal C}_{\ell}^{TE}\right)^2\frac{\partial {\cal C}_{\ell}^{TT}}
{\partial R}\frac{\partial {\cal C}_{\ell}^{EE}}{\partial R}
\nonumber\\
 &~&
 \left.+4{\cal C}_{\ell}^{TE}
 \left( {\cal C}_{\ell}^{TT}\frac{\partial {\cal C}_{\ell}^{EE}}{\partial R}+
{\cal C}_{\ell}^{EE}\frac{\partial {\cal C}_{\ell}^{TT}}{\partial R}
\right)\frac{\partial {\cal C}_{\ell}^{TE}}{\partial R}\right\}
+\frac{R^2(2\ell+1)f_{\rm sky}}{2({\cal C}_{\ell}^{BB})^2}\frac{\partial 
{\cal C}_{\ell}^{BB}}{\partial R}\frac{\partial {\cal C}_{\ell}^{BB}}{\partial R}.
\end{eqnarray}
The natural break-up of Eq.~(\ref{s3}) into combinations of the 
information channels is $TT+TE+EE+BB$, $TT+TE+EE$ and $BB$ alone.

\subsubsection{Signal to noise ratio for different combinations of information 
channels
\label{section4.2.1}}

\begin{figure}
\centerline{\includegraphics[width=18cm,height=15cm]{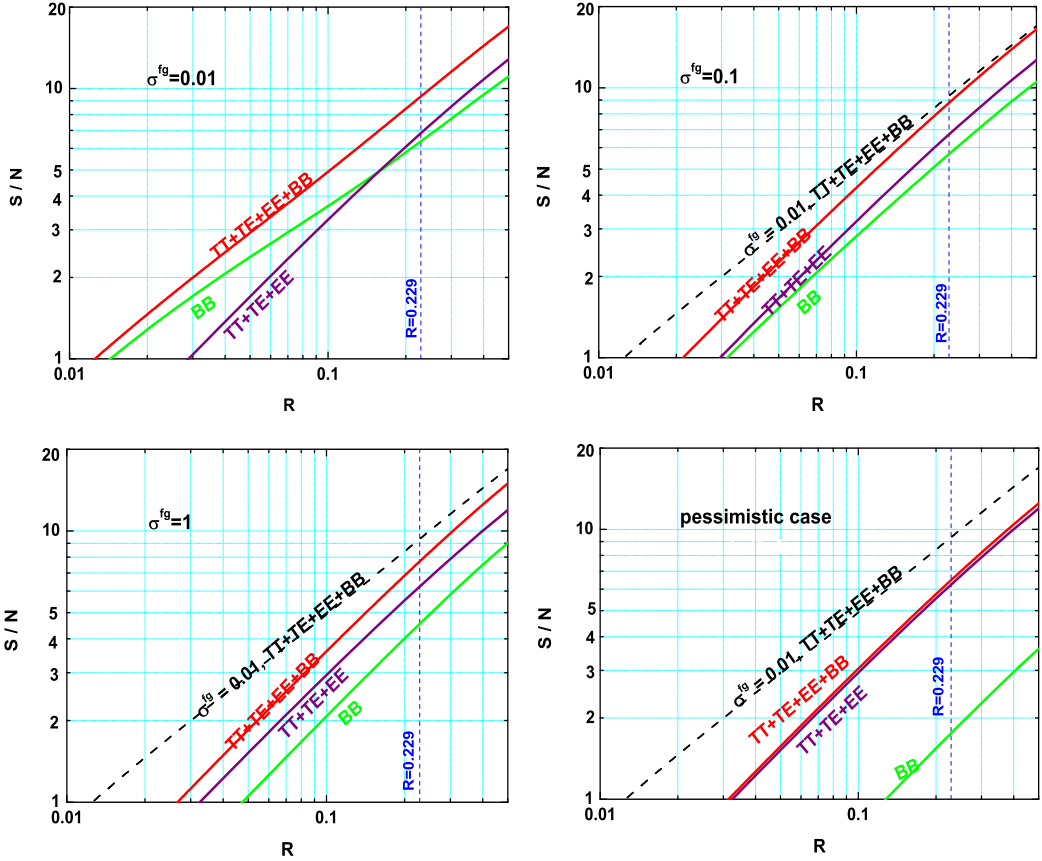}}
\caption{The $S/N$ for different combinations of the information channels, 
$TT+TE+EE+BB$, $TT+TE+EE$ and $BB$.}\label{figurev11}
\end{figure}

Using either all terms in Eq.~(\ref{s3}), or everything without $BB$, or
$BB$ alone, we calculate $(S/N)^2$, Eq.~(\ref{s2}), for three combinations 
of channels:  $TT+TE+EE+BB$, $TT+TE+EE$ and $BB$ alone. The $(S/N)^2$ for 
the first (full) combination is the sum of $(S/N)^2$ for the other two. Since 
the foreground removal is a major concern, we separate the results into four 
groups - $\sigma^{\rm fg} = 0.01,~0.1,~1$ and the pessimistic case. The 
results for $S/N$ are shown in Fig.~\ref{figurev11} in four panels. 
Certainly, the (red) lines marked by $TT+TE+EE+BB$ in four panels are the 
same lines that are shown in the left panel of Fig.~\ref{figurev2} for the 
corresponding case. The copy of the line representing the optimistic scenario 
in the upper left panel, i.e. $TT+TE+EE+BB$ together with 
$\sigma^{\rm fg} = 0.01$, is shown by a (black) dashed line in other panels 
as a reminder of what can be achieved in the optimistic scenario.

Surely, the combination $TT+TE+EE+BB$ is more sensitive than any of the
other two, $TT+TE+EE$ and $BB$ alone. For example, in the case 
$\sigma^{\rm fg}=0.1$, the use of all channels provides $S/N$ which is 
$\sim50\%$ greater than $BB$ alone and $\sim30\%$ greater than $TT+TE+EE$. 
The situation is even more peculiar in the pessimistic case. The ML model 
(\ref{best-fit}) can be barely seen through the popular $B$-modes alone, 
because the $BB$ channel gives only $S/N=1.75$, whereas the use of all 
channels can provide a confident detection with $S/N=6.48$. 

Comparing $TT+TE+EE$ with $BB$, one can see that the first method is better 
than the second, except in the case when $\sigma^{\rm fg}=0.01$ and $R$ is 
small ($R<0.16$). In the pessimistic case, the role of the $BB$ channel is so
small that the $TT+TE+EE$ method provides essentially the same sensitivity as 
all channels $TT+TE+EE+BB$ together.

Considering the $S/N$ for $BB$ alone, it is worth noting that since the $BB$ 
channel is not sensitive to $A_s$ and $n_s$ the uncertainty (\ref{deltaR2}) 
reduces to (\ref{deltaR1}). Therefore, although the results for $BB$ channel 
alone, shown in Fig.~\ref{figurev11}, were derived under the assumption of a 
single unknown parameter $R$, they apply also in the general case when all 
three parameters, $R$, $n_s$, $A_s$, are supposed to be determined from the 
temperature and polarization data.

\subsubsection{Multipole decomposition of $S/N$ \label{section4.2.2}}

It is seen from Eq.~(\ref{s2}) the the total $(S/N)^2$ is a sum 
over $\ell$-contributions $(S/N)_{\ell}^2$ given by Eq.~(\ref{s3}). 
Formally, the sum can extend to large $\ell$'s, but a smaller $R$-signal and
larger noises make the large $\ell$'s irrelevant anyway. We go up to 
$\ell_{max} =100$.

\begin{figure}
\centerline{\includegraphics[width=18cm,height=15cm]{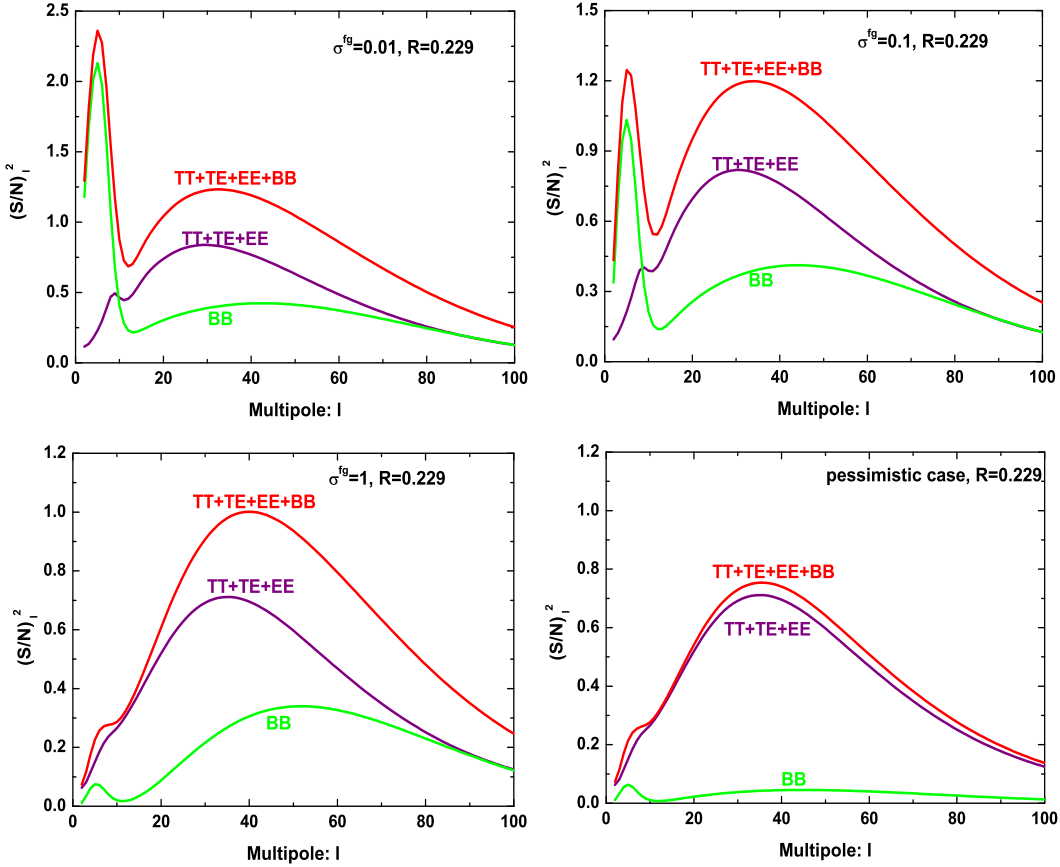}}
\caption{The individual $(S/N)_{\ell}^{2}$ as functions of $\ell$ for 
different combinations of information channels and different
levels of foreground contamination. Calculations are done for the ML
model (\ref{best-fit}) with $R=0.229$.}\label{figurep41}
\end{figure}

\begin{figure}
\centerline{\includegraphics[width=18cm,height=15cm]{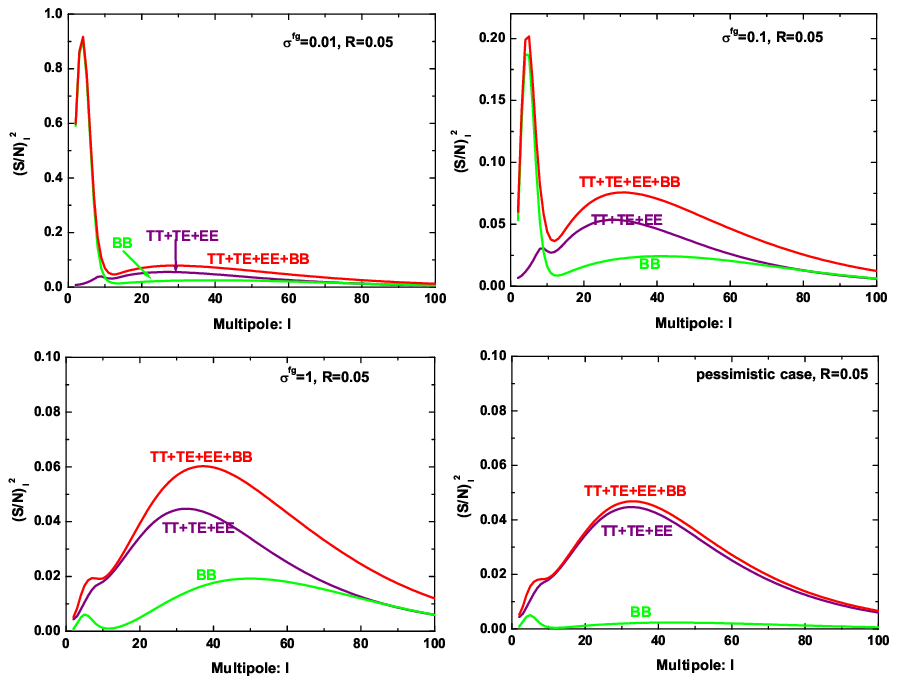}}
\caption{The same illustrations as in Fig.~\ref{figurep41}, but for the 
model (\ref{1Dmodel}) with $R=0.05$.}\label{figurep42}
\end{figure}

In Fig.~\ref{figurep41} and Fig.~\ref{figurep42} we plot the quantity 
$(S/N)_{\ell}^2$ as a function of $\ell$ for different combinations of 
information channels and different $\sigma^{\rm fg}$, including the 
pessimistic case. Fig.~\ref{figurep41} presents the multipole 
decomposition for the ML model (\ref{best-fit}) with $R=0.229$, while 
Fig.~\ref{figurep42} sorts out the $R=0.05$ model characterized by $S/N =3$ 
for $\sigma^{\rm fg}=0.01$ (see left panel in Fig.~\ref{figurev2}). Note the 
differences in scaling on vertical axes in Fig.~\ref{figurep41} and 
Fig.~\ref{figurep42}. 

Both figures show again that a very deep foreground cleaning, 
$\sigma^{\rm fg}=0.01$, makes the very low (reionization) multipoles 
$\ell\simeq 10$ the major contributors to the total $(S/N)^2$, and mostly from
the $BB$ channel. This is especially true for the lower-$R$ model $R=0.05$.
However, for large $\sigma^{\rm fg}=0.1,~1$, and especially in the 
pessimistic case (see the lower right panels in Figs. \ref{figurep41} 
and \ref{figurep42}), the role of the $BB$ channel becomes very small 
at all $\ell$'s. These detailed illustrations in terms of multipole 
decomposition of $(S/N)^2$ are of course fully consistent with the integrated 
results of Sec.~\ref{section4.2.1}.

At the same time, as Figs. \ref{figurep41} and \ref{figurep42} illustrate, the 
$\ell$-decomposition of $(S/N)^2$ for $TT+TE+EE$ combination depends only 
weakly on the level of $\sigma^{\rm fg}$. Furthermore, in this method, 
the signal-to-noise curves generally peak at $\ell\sim(20-50)$. Therefore, it 
will be particularly important for {\it Planck} mission to avoid excessive 
noises in this region of multipoles.


\subsection{On the possibilities to get a better value for $S/N$ 
\label{section4.3}}

So far, the $S/N$ was calculated using the generally quoted realistic 
assumptions about instrumental and environmental noises. It appears that some 
reserves to get better values for $S/N$ still exist. These possibilities
seem speculative but worth exploring. First, one may find a way of using 
the outputs of all seven frequency channels (listed in Table \ref{table1}) 
for the CMB analysis. The summation over $i=1,\cdot\cdot\cdot,7$, instead of $i=1,2,3$, 
would effectively reduce the instrumental part of noise in 
$N_{\ell}^{XX}$ (\ref{totalnoise}). Second, we may be lucky and the 
Dust B model turns out to be correct one, rather than the more severe 
Dust A model. This would reduce the foreground part of noise in 
$N_{\ell}^{XX}$ (\ref{totalnoise}).

\begin{figure}
\centerline{\includegraphics[width=20cm,height=14cm]{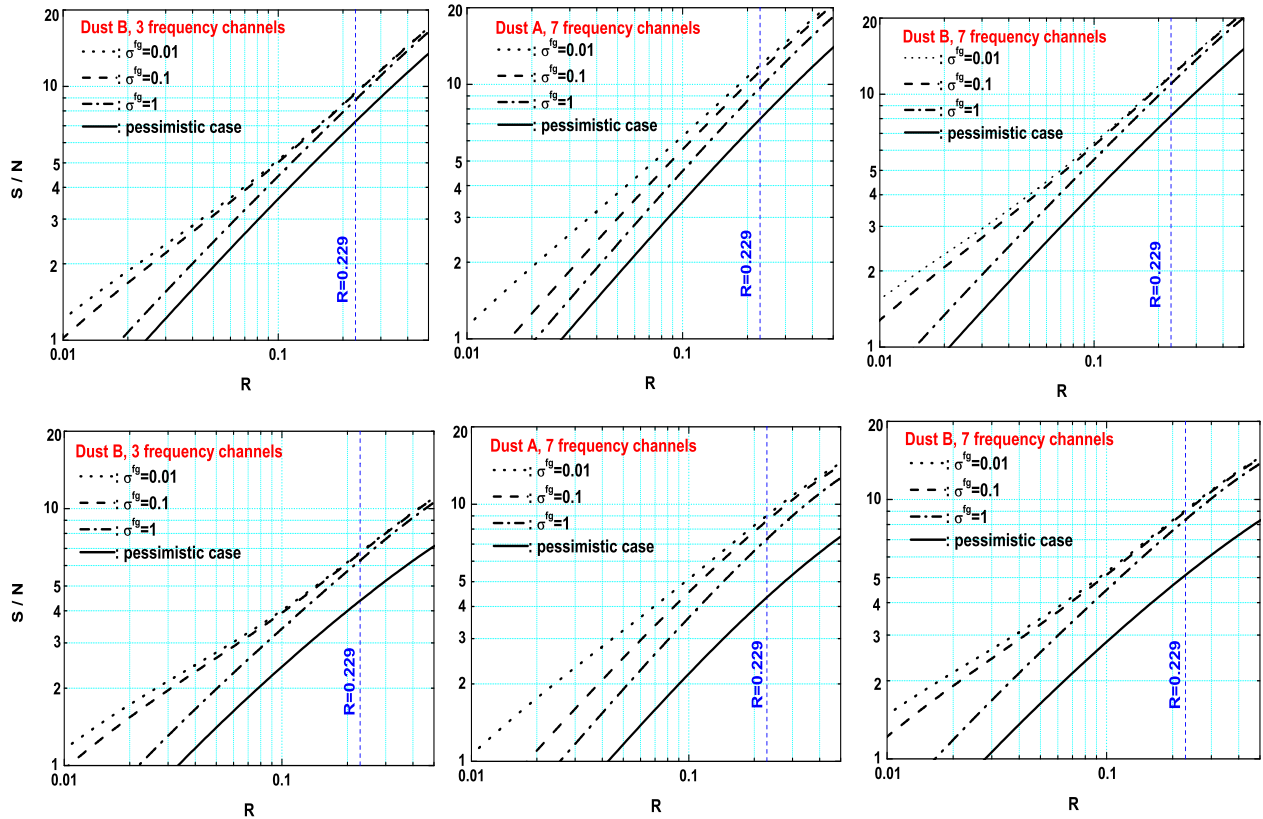}}
\caption{The improved values of $S/N$. The three upper panels should be 
compared with the left panel in Fig.~\ref{figurev2}, and the three lower 
panels with the right panel in Fig.~\ref{figurev2}.} 
\label{figurev44}
\end{figure}

Below, we consider three possibilities of these improvements in $S/N$. Namely, 
the use of 3 frequency channels and the validity of Dust B model, the use of 7 
frequency channels and the validity of Dust A model, and the `dream case' of
using all 7 frequency channels in the conditions of validity of Dust B model.
For these three options, we do exactly the same calculations that were done 
in Sec.~\ref{section4.1.2} and depicted in Fig.~\ref{figurev2}. The left 
panel in Fig.~\ref{figurev2} translates into the three upper panels in 
Fig.~\ref{figurev44}, and the right panel in Fig.~\ref{figurev2} translates 
into the three lower panels in Fig.~\ref{figurev44}.

Certainly, one sees considerable improvements in $S/N$, especially in the
`dream case' scenario (upper right and lower right panels in 
Fig.~\ref{figurev44}). For example, concentrating on the solid line in the 
lower right panel, one finds that the ML model (\ref{best-fit}) with 
$R=0.229$ would be detectable at the level $S/N = 5.11$, instead of 
$S/N = 3.65$ that we found in the right panel in Fig.~\ref{figurev2}. 
This would significantly increase the chance of observing relic gravitational 
waves with the help of the {\it Planck} mission.

\section{Conclusions}

Being in the possession of general theoretical confidence that the relic 
gravitational waves are expected to be present in the observed CMB 
anisotropies, we improved our previous analysis of WMAP5 data and made 
comprehensive forecasts for the {\it Planck} mission. 

The improvements in the analysis of the WMAP5 $TT$ and $TE$ observations 
made our approach more general and stable. The new analysis avoids 
phenomenological relations and restrictions, and deals directly with 
the complete 3-parameter likelihood function and its marginalizations. The 
result of this analysis is very close to the previous one: the maximum 
likelihood value for the quadrupole ratio $R$ is $R=0.229$. This means 
that approximately 20\% of the temperature quadrupole is caused by 
gravitational waves and 80\% by density perturbations. Although the 
uncertainties due to large noises are still large, this result can be 
regarded as an indication of the presence of relic gravitational waves 
in the lower $\ell$ CMB anisotropies (we would love to call it a suspected 
detection, but we resist this temptation). 

We identify and study in detail the reasons by which the contribution of 
relic gravitational waves can be overlooked in a data analysis. One of 
the reasons is the unjustified reliance on constancy of the 
spectral index $n_s$ and the inclusion of data from very high 
multipoles $\ell$. Another reason - a too narrow understanding of the 
problem as the search for $B$-modes of polarization, rather than the 
detection of relic gravitational wave with the help of all correlation 
functions. 

Our forecasts for {\it Planck} are also based on our analysis of WMAP5 
observations. We identify the whole family of models, that is, the sets of 
perturbation parameters $R$, $n_s$, $A_s$, which are almost as good as the 
unique model with the maximum likelihood, $\mathcal{L} = 1$, set of parameters 
$R=0.229, n_s=1.086, A_s=1.920\times10^{-9}$. For the same observational 
data, these sets of parameters return reasonably high numerical values of 
the likelihood function $\mathcal{L}$. We formulate our forecasts for this
family of models, which is characterized by observationally preferred sets 
of parameters, rather than choosing the models and parameters at random and 
blindly. Our forecasts, based on the Fisher matrix techniques, refer to 
achievable signal-to-noise ratios $S/N$. We analyze many sources of noise 
and explore various `optimistic', `pessimistic' and `dream case' scenarios. 
We discuss the circumstances in which the $B$-mode detection is very 
inconclusive, at the level $S/N =1.75$, whereas a smarter data analysis 
reveals the same gravitational wave signal at $S/N= 6.48$. 

The very encouraging final result is that, even under unfavourable conditions
in terms of instrumental noises and foregrounds, the relic gravitational 
waves, if they are characterized by the WMAP5 maximum likelihood value 
$R=0.229$, will be detected by {\it Planck} at the level $S/N = 3.65$.  


{\bf Acknowledgements}

Wen Zhao is partially supported by Chinese NSF Grants No. 10703005 and 
No. 10775119.


\appendix

\section{Fisher information matrix in the CMB analysis \label{appa}}

In the CMB data analysis, the general form of the likelihood function, up 
to a normalization constant, is
\begin{eqnarray}
\mathcal{L}\propto \prod_{\ell}
f(D_{\ell}^{TT},D_{\ell}^{TE},D_{\ell}^{EE},D_{\ell}^{BB}).
\end{eqnarray}
Since $D_{\ell}^{BB}$ is independent of the rest of variables 
$D_{\ell}^{TT}$, $D_{\ell}^{TE}$, $D_{\ell}^{EE}$ \cite{zbg}, the likelihood 
function factorizes,
\begin{eqnarray}\label{ctelikelihood2}
\mathcal{L}\propto \prod_{\ell}
f(D_{\ell}^{TT},D_{\ell}^{TE},D_{\ell}^{EE})f(D_{\ell}^{BB}).
\end{eqnarray}
The pdf $f(D_{\ell}^{TT},D_{\ell}^{TE},D_{\ell}^{EE})$ is the Wishart 
distribution given by Eq.~(\ref{wishart}), whereas $f(D_{\ell}^{BB})$ is 
the $\chi^2$ distribution:
\begin{eqnarray}
f(D_{\ell}^{BB})=\frac{nV^{(n-2)/2}e^{-V/2}}{2^{n/2}
\Gamma(n/2)(C_{\ell}^{BB}+N_{\ell}^{BB})},
\end{eqnarray}
where $V$ is 
$V\equiv n(D_{\ell}^{BB}+N_{\ell}^{BB})/(C_{\ell}^{BB}+N_{\ell}^{BB})$.
The likelihood function is a function of the sought after parameters $s_i$, 
which in our case are perturbation parameters $A_t$, $A_s$, $n_t$, $n_s$. 
They enter the likelihood function through their presence in the power spectra 
$C_{\ell}^{XX'}$. The $BB$ part of $\mathcal{L}$ depends only on $A_t$ and
$n_t$. We have reduced the set of parameters $s_i$ to $R$, $n_s$, $A_s$. 

The Fisher information matrix is a measure of the width and shape
of the likelihood function, as a function of the parameters, 
around its maximum. The Fisher matrix formalism is used for estimation of
the accuracy with which the parameters of interest can be found from the data 
\cite{fisher1,fisher2}. The elements of the matrix are expectation values 
of the second derivatives of logarithm of the likelihood function with 
respect to $s_i$, 
 \begin{eqnarray}\label{fisherdefination}
 F_{s_is_j}=\left.\left\langle-\frac{\partial^2{(\ln\cal{L}})}{\partial s_i\partial
 s_j}\right\rangle\right|_{{s}_i=\bar{s}_i},
 \end{eqnarray}
where $\bar{s}_i$ are the true values of ${s}_i$ (i.e.~values where the average of the first derivative of $\ln \cal{L}$ vanishes). 
The angle brackets denote the integration over the joint pdf for 
all $D_{\ell}^{XX'}$. 

Inserting (\ref{ctelikelihood2}) into (\ref{fisherdefination}), one can 
calculate the Fisher matrix, 
\begin{eqnarray}
\label{fishermatrix}
 F_{s_{i}s_{j}}=\sum_{\ell}\sum_{XX',YY'}\frac{\partial {C}_{\ell}^{XX'}}
{\partial s_{i}}{{\rm Cov}^{-1} (D_{\ell}^{XX'},D_{\ell}^{YY'})}
\frac{\partial {C}_{\ell}^{YY'}}{\partial s_{j}},
\end{eqnarray}
where ${\rm Cov}^{-1}$ is the inverse of the covariance matrix. (Result 
(\ref{fishermatrix}) coincides with Eq.~(7) in \cite{fisher2}.) The 
non-vanishing components of the covariance matrix are given by
 \begin{eqnarray} 
 {\rm Cov} (D_{\ell}^{XX},D_{\ell}^{XX})&=&\frac{2}{(2\ell+1)f_{\rm sky}}
(C_{\ell}^{XX}+N_{\ell}^{XX})^2~~~(X=T,E,B),\nonumber\\
 {\rm Cov} (D_{\ell}^{TE},D_{\ell}^{TE})&=&\frac{1}{(2\ell+1)f_{\rm sky}}
[(C_{\ell}^{TE})^2+(C_{\ell}^{TT}+N_{\ell}^{TT})(C_{\ell}^{EE}+N_{\ell}^{EE})],
\nonumber\\
 {\rm Cov} (D_{\ell}^{TT},D_{\ell}^{EE})&=&\frac{2}{(2\ell+1)f_{\rm sky}}
(C_{\ell}^{TE})^2,\nonumber\\
 {\rm Cov} (D_{\ell}^{TT},D_{\ell}^{TE})&=&\frac{2}{(2\ell+1)f_{\rm sky}}
C_{\ell}^{TE}(C_{\ell}^{TT}+N_{\ell}^{TT}),\nonumber\\
 {\rm Cov} (D_{\ell}^{EE},D_{\ell}^{TE})&=&\frac{2}{(2\ell+1)f_{\rm sky}}
C_{\ell}^{TE}(C_{\ell}^{EE}+N_{\ell}^{EE}).\nonumber
 \end{eqnarray}

When a particular parameter $s$ is estimated from the data, 
while other parameters are assumed to be known, the uncertainty in the 
determination of this parameter is given by $\Delta s=1/\sqrt{F_{ss}}$. 
However, if all parameters are estimated from the data, the uncertainty 
rises to $\Delta s=\sqrt{(F^{-1})_{ss}}$. The second uncertainty is always
larger than the first one or equal to it. 

In this work, we are mostly interested in the parameter $R$, which quantifies 
the contribution of relic gravitational waves to the CMB anisotropies. 
The definition of the signal-to-noise ratio $S/N$ in Eq.~(\ref{snr}) requires 
the specification of the uncertainty $\Delta R$. If other parameters are fixed 
and only $R$ is derived from the data, this uncertainty is given by the
matrix element $F_{RR}$, 
\begin{eqnarray} 
\label{deltaR1}
\Delta R={1}/{\sqrt{F_{ RR}}}.
\end{eqnarray}
Explicit expression for $F_{RR}$ follows from Eq.~(\ref{fishermatrix}),
\begin{eqnarray}
\label{FRR}
 F_{RR}&=&\sum_{\ell}\frac{R^2(2\ell+1)f_{\rm sky}}{2\left({\cal C}_{\ell}^{TT}
{\cal C}_{\ell}^{EE}-({\cal C}_{\ell}^{TE})^2\right)^2}\times\left\{({\cal C}_{\ell}^{TT})^2
\frac{\partial {\cal C}_{\ell}^{EE}}{\partial R}\frac{\partial {\cal C}_{\ell}^{EE}}
{\partial R} +({\cal C}_{\ell}^{EE})^2\frac{\partial {\cal C}_{\ell}^{TT}}
{\partial R}\frac{\partial {\cal C}_{\ell}^{TT}}{\partial R}\right.\nonumber\\
 &~&
 +2\left({\cal C}_{\ell}^{TT}{\cal C}_{\ell}^{EE}+\left({\cal C}_{\ell}^{TE}\right)^2\right)
 \frac{\partial {\cal C}_{\ell}^{TE}}{\partial R}\frac{\partial {\cal C}_{\ell}^{TE}}{\partial R} 
+2\left({\cal C}_{\ell}^{TE}\right)^2\frac{\partial {\cal C}_{\ell}^{TT}}
{\partial R}\frac{\partial {\cal C}_{\ell}^{EE}}{\partial R}
\nonumber\\
 &~&
 \left.+4{\cal C}_{\ell}^{TE}
 \left( {\cal C}_{\ell}^{TT}\frac{\partial {\cal C}_{\ell}^{EE}}{\partial R}+
{\cal C}_{\ell}^{EE}\frac{\partial {\cal C}_{\ell}^{TT}}{\partial R}
\right)\frac{\partial {\cal C}_{\ell}^{TE}}{\partial R}\right\}
+\frac{R^2(2\ell+1)f_{\rm sky}}{2({\cal C}_{\ell}^{BB})^2}\frac{\partial 
{\cal C}_{\ell}^{BB}}{\partial R}\frac{\partial {\cal C}_{\ell}^{BB}}{\partial R},
\end{eqnarray}
where $\mathcal{C}_{\ell}^{XX}\equiv C_{\ell}^{XX}+N_{\ell}^{XX}$, and 
$\mathcal{C}_{\ell}^{TE}\equiv C_{\ell}^{TE}$.

If other parameters are being determined together with $R$, the 
uncertainty $\Delta R$ is given by the $RR$ element of the inverse matrix, 
\begin{eqnarray}
\label{deltaR2}
\Delta R={\sqrt{\left(F^{-1}\right)_{RR}}}.
\end{eqnarray}
The uncertainty (\ref{deltaR2}) is always greater than (\ref{deltaR1}) or 
equal to it. We use (\ref{deltaR1}) and (\ref{deltaR2}) for numerical 
evaluation of $S/N$.


\section{The instrumental and environmental noise power spectra\label{appb}}

CMB experiments use several frequency channels (distinguished by the 
label $i$) which have specific instrumental and environmental noises at each
frequency $\nu_i$. The optimal combination of the channels gives the total 
effective noise power spectrum $N_{\ell}^{XX}$ \cite{foreground3,cmbpol,Bowden2004}
\begin{eqnarray}
\label{totalnoise}
 [N_{\ell}^{XX}]^{-2}=\sum_{i\ge j}\left[(N_{{\rm fg},\ell}^{XX}(i)+
N_{{\rm ins},\ell}^{XX}(i)) (N_{{\rm fg},\ell}^{XX}(j)+
N_{{\rm ins},\ell}^{XX}(j))\frac{1}{2}(1+\delta_{ij})\right]^{-1},
\end{eqnarray}
where $N_{{\rm ins},\ell}^{XX}(i)$ and $N_{{\rm fg},\ell}^{XX}(i)$ are the 
instrumental and residual foreground noise power spectra, respectively.
The total effective noise power spectrum $N_{\ell}^{XX}$ enters the Fisher 
matrix (\ref{fishermatrix}).  

In the evaluation of noise power spectra we set the window function equal 
to 1, which is a good approximation for the multipoles considered in this 
paper, $\ell\leq 220$. The instrumental characteristics of the {\it Planck's}  
frequency channels are reported in Table \ref{table1} based 
on \cite{planck}. This Table provides $N_{{\rm ins},\ell}^{XX}(i)$. As for the 
polarized foregrounds ($XX=EE,BB$), we focus on the synchrotron (S) and 
dust (D) emissions. The foreground contamination 
is quantified by the parameter $\sigma^{{\rm fg}} = 1,~0.1,~0.01$ which 
multiplies the power spectra $C_{S,\ell}^{XX}(i)$, $C_{D,\ell}^{XX}(i)$ of
the accepted foreground models. The smaller $\sigma^{{\rm fg}}$ the deeper
cleaning. The residual foreground noise is given by 
\cite{foreground3}
\begin{eqnarray}
\label{fore}
N_{{\rm fg},\ell}^{XX}(i)=\sum_{{f}=S,D}\left[C_{{f},\ell}^{XX}(i)
\sigma^{{\rm fg}}+\mathcal{N}_{{f},\ell}^{XX}(i)\right],
\end{eqnarray}
where $\mathcal{N}_{f,\ell}^{XX}(i)$ is the noise power spectrum arising from 
the cleaning procedure itself in the presence of instrumental noise.

\begin{table}
\caption{Summary of {\it Planck} Instrumental Characteristics }
\begin{center}
\label{table1}
\begin{tabular}{|c|c|c|c|c|c|c|c|}
  \hline
 \multicolumn{1}{|c|}{Instrument Characteristic}  & \multicolumn{3}{c|}{LFI}& \multicolumn{4}{c|}{HFI}\\
    \hline
    Center Frequency~[GHz] & ~~30~~ & ~~44~~ & ~~70~~ & ~100~ & ~143~ & ~217~ & ~353~  \\
    \hline
    Angular Resolution~[FWHM~arcmin] & ~~33~~ & ~~24~~ & ~~14~~ & ~10~ & ~7.1~ & ~5.0~ & ~5.0~  \\
       \hline
    $N_{{\rm ins},\ell}^{TT}$~[$10^{-4}\mu$K$^2$] & ~~27.37~~ & ~~26.38~~ & ~~27.20~~ & ~3.93~ & ~1.53~ & ~3.62~ & ~33.94~  \\
       \hline
    $N_{{\rm ins},\ell}^{EE}$ and $N_{{\rm ins},\ell}^{BB}$~[$10^{-4}\mu$K$^2$] & ~~53.65~~ & ~~55.05~~ & ~~55.28~~ & ~10.05~ & ~5.58~ & ~15.09~ & ~139.50~  \\
  \hline
\end{tabular}
\end{center}
\end{table}

Following \cite{foreground3,cmbpol,foreground2}, for the $\ell$ and $\nu_i$ 
dependences of the synchrotron and dust emissions we take 
\begin{eqnarray}
\label{synchr}
 C_{S,\ell}^{XX}(i)=A_S\left(\frac{\nu_i}{\nu_0}\right)^{2\alpha_S}
\left(\frac{\ell}{\ell_0}\right)^{\beta_S}
\end{eqnarray}
and
 \begin{eqnarray}
\label{dust}
 C_{D,\ell}^{XX}(i)=p^2A_D\left(\frac{\nu_i}{\nu_0}\right)^{2\alpha_D}
\left(\frac{\ell}{\ell_0}\right)^{\beta_D^{XX}}
 \left[\frac{e^{h\nu_0/kT}-1}{e^{h\nu_i/kT}-1}\right]^2.
\end{eqnarray}
In Eq.~(\ref{dust}), $p$ is the dust polarization fraction, estimated to be 
$5\%$ \cite{foreground3}, and $T$ is the temperature of the dust grains, 
assumed to be constant across the sky $T=18$K \cite{foreground3}. Other 
parameters in Eqs.~(\ref{synchr}), (\ref{dust}) are specified in Table 
\ref{table2} taken from \cite{cmbpol}.

\begin{table}
\caption{Assumptions about foregrounds \cite{cmbpol} }
\begin{center}
\label{table2}
\begin{tabular}{|c|c|c|c|}
    \hline
    Parameter &~~~{Synchrotron}~~~&~~~~~~~~~~{Dust A}~~~~~~~~~~&~~~{Dust B}~~~   \\
    \hline
    $A_{S,D}$  & $4.7\times 10^{-5}$~$\mu$K$^{2}$ & $1.0$~$\mu$K$^{2}$ & $1.2\times10^{-4}$~$\mu$K$^{2}$   \\
    \hline
    $\nu_0$   & 30~GHz & 94~GHz & 94~GHz   \\
    \hline
    $\ell_0$   & 350 & 10 & 900   \\
    \hline
    $\alpha$   & -3 & 2.2 & 2.2   \\
    \hline
    $\beta^{EE}$   & -2.6 & -2.5 & -1.3   \\
    \hline
    $\beta^{BB}$   & -2.6 & -2.5 & -1.4   \\
  \hline
\end{tabular}
\end{center}
\end{table}

The noise term $\mathcal{N}_{f,\ell}^{XX}(i)$ ($f=S,D$) entering 
Eq.~(\ref{fore}) was calculated in \cite{foreground3,cmbpol} 
\[
\mathcal{N}_{f,\ell}^{XX}(i)=\frac{N_{{\rm ins},\ell}^{XX}(i)}{n_{\rm chan}
(n_{\rm chan}-1)/4}\left(\frac{\nu_i}{\nu_{\rm ref}}\right)^{2\alpha}.
\]
Here, $n_{\rm chan}$ is the total number of frequency channels used in making 
the foreground template map, and $\nu_{\rm ref}$ is the frequency of the
reference channel. In the case of the dust, $\nu_{\rm ref}$ is the highest 
frequency channel included in the template making, while in the case of the 
synchrotron, $\nu_{\rm ref}$ is the lowest frequency channel. The value of 
$\alpha$ is given in Table \ref{table2} for different foreground models.

All components of noise are used in numerical calculations of the total noise 
(\ref{totalnoise}).



\end{document}